\documentclass[aps,pra,twocolumn,showpacs,groupedaddress,amsmath]{revtex4-1}

\usepackage{bm}
\usepackage[english]{babel}
\usepackage{graphicx}
\usepackage{units}
\usepackage{hyperref}
\usepackage{ulem}

\usepackage{color}

\begin{document}

\title{Efficient production of long-lived ultracold $\mathrm{\mathbf{Sr}}_{\mathbf{2}}$ molecules}

\author{Alessio Ciamei}
\email[]{Sr2molecules@strontiumBEC.com}
\author{Alex Bayerle}
\author{Chun-Chia Chen}
\author{Benjamin Pasquiou}
\author{Florian Schreck}
\affiliation{Van  der  Waals-Zeeman  Institute,  Institute  of  Physics,  University  of Amsterdam, Science  Park  904,  1098  XH  Amsterdam,  The  Netherlands}

\date{\today}



\begin{abstract}
We associate Sr atom pairs on sites of a Mott insulator optically and coherently into weakly-bound ground-state molecules, achieving an efficiency above 80\%. This efficiency is 2.5 times higher than in our previous work [S. Stellmer, B. Pasquiou, R. Grimm, and F. Schreck, \href{https://doi.org/10.1103/PhysRevLett.109.115302}{Phys. Rev. Lett. \textbf{109}, 115302 (2012)}] and obtained through two improvements. First, the lifetime of the molecules is increased beyond one minute by using an optical lattice wavelength that is further detuned from molecular transitions. Second, we compensate undesired dynamic light shifts that occur during the stimulated Raman adiabatic passage (STIRAP) used for molecule association. We also characterize and model STIRAP, providing insights into its limitations. Our work shows that significant molecule association efficiencies can be achieved even for atomic species or mixtures that lack Feshbach resonances suitable for magnetoassociation.
\end{abstract}

\maketitle

\section{Introduction}
\label{sec:Introduction}

Over the last fifteen years considerable experimental effort has been invested into the realization of ultracold molecular samples. Ultracold molecules hold promise for unveiling novel phases of matter near quantum degeneracy, implementing quantum information protocols, and enabling precision measurements beyond atomic physics \cite{Krems2009cmc,Carr2009ReviewMolecules,Baranov2012cmt}. Ultracold dimers in their rovibrational ground state can be created in a two-step process from ultracold atoms. In the first step, atom pairs are associated into weakly-bound molecules and in the second step, the molecules are transferred from a weakly-bound state to the rovibrational ground state \cite{Ni2008HighPSDMol, Danzl2008BoundMolStirap}. So far the first step in these experiments has relied on the existence of magnetically tuneable Feshbach resonances \cite{Kohler2006RevFeshbachMol, Ospelkaus2006FeshbachMolLattice}. By ramping an external magnetic field adiabatically across such a resonance, a coherent transfer between a pair of free atoms and a molecular bound state can be accomplished. Such magneto-association is hard or even impossible for a vast class of atomic systems of interest, for instance combinations of an alkali metal and an alkaline-earth metal or pairs of alkaline-earth metal atoms. The former systems possess only extremely narrow magnetic Feshbach resonances \cite{Zuchowski2010RbSrFeshbach,Brue2013pof}, the latter none at all.

Production of ultracold weakly-bound ground-state $\mathrm{Sr}_{2}$ molecules was achieved in our previous work \cite{Stellmer2012Sr2Mol} and in \cite{Reinaudi2012Sr2Mol}, relying respectively on coherent and non-coherent optical transfer schemes, thus overcoming the absence of magnetic Feshbach resonances in the non-magnetic ground state of these atoms. More recently, two-photon coherent transfer of cold Rb atom pairs into ground-state Rb$_2$ molecules was demonstrated using a frequency-chirped laser pulse \cite{PhysRevLett.115.173003}. The coherent population transfer of \cite{Stellmer2012Sr2Mol} was stimulated Raman adiabatic passage (STIRAP), which evolves a dark state from a pair of atoms into a molecule \cite{Gaubatz1990FirstSTIRAP, Bergmann2015RevSTIRAP}. Unfortunately, because of losses by non-adiabatic coupling and short lifetime of the molecules, the molecule association efficiency was only 30\%, far below the efficiency potentially achievable by STIRAP. Moreover, the short lifetime hindered further usage of the molecular sample.

In this article we show how to overcome these limitations. As in \cite{Stellmer2012Sr2Mol} we investigate the production of $^{84}\mathrm{Sr}_{2}$ ultracold ground-state molecules by STIRAP starting from a Mott insulator (MI). We increase the lifetime of the molecules to over one minute by using an optical lattice wavelength that, unlike before, is far detuned from any molecular transition. We identify that the resulting STIRAP efficiency of slightly above $\unit[50]{\%}$ is limited by the finite lifetime of the dark state arising from unwanted light shifts. We show how to overcome these light shifts with the help of an additional compensation beam \cite{Yatsenko1997STIRAPwithStarkShiftComp}, leading to a STIRAP efficiency above $\unit[80]{\%}$. Our work validates a general way of producing large samples of weakly-bound molecules without relying on Feshbach resonances. This will open the path for new classes of ultracold dimers useful for metrology experiments \cite{Hudson2006FineStrucVar, Kajita2014ProtToElec, Flambaum2007MassProtElecVar}, for ultracold chemistry \cite{Krems2008RevColdChemistry, Miranda2011StereodynamicMolKRb} and for quantum simulation experiments relying on a strong permanent electric dipole moment \cite{Micheli2006ToolboxPolMol, Barnett2006QPhaseDiploarMol, Buchler2007QPhase2DPolarMol}.

This article is organized as follows. In Sec.~\ref{sec:Overview} we present an overview of our experimental strategy. In Sec.~\ref{sec:Model and parameters} we introduce the model used to describe the STIRAP and we discuss the constraints imposed on the relevant experimental parameters when high transfer efficiency is required. In Sec.~\ref{sec:Ingredients} we describe the experimental sequence leading to the initial atomic sample and the optical scheme for the creation of photoassociation (PA) laser light. In Sec.~\ref{subsec:MI Parameters}, we measure relevant parameters of the system and we show the effect of the lattice on the free-bound Rabi frequency. In Sec.~\ref{subsec:MI STIRAP} we use STIRAP to associate atom pairs into molecules, achieving a transfer efficiency of $\sim \unit[50]{\%}$, and we show that this process is limited by the finite lifetime of the dark state, arising from unwanted light shifts on the binding energy of the ground-state molecule. This challenge is overcome in Sec.~\ref{subsec:Light-shift compensation} by modifying the standard scheme with the addition of a light-shift compensation beam, reaching an efficiency higher than $\unit[80]{\%}$. Finally, in Sec.~\ref{subsec:Sample characterization} we characterize the effects of the lattice light shifts on STIRAP and we measure the molecule lifetime.

\section{Experimental strategy}
\label{sec:Overview}

We will now discuss how one can optically and coherently associate pairs of atoms into ground-state molecules. We will explain how STIRAP works and can reach near-unit molecule association efficiency. We motivate the use of a MI as initial atomic sample and specify the STIRAP implementation used here.

\begin{figure}[t]
\includegraphics[width=\columnwidth]{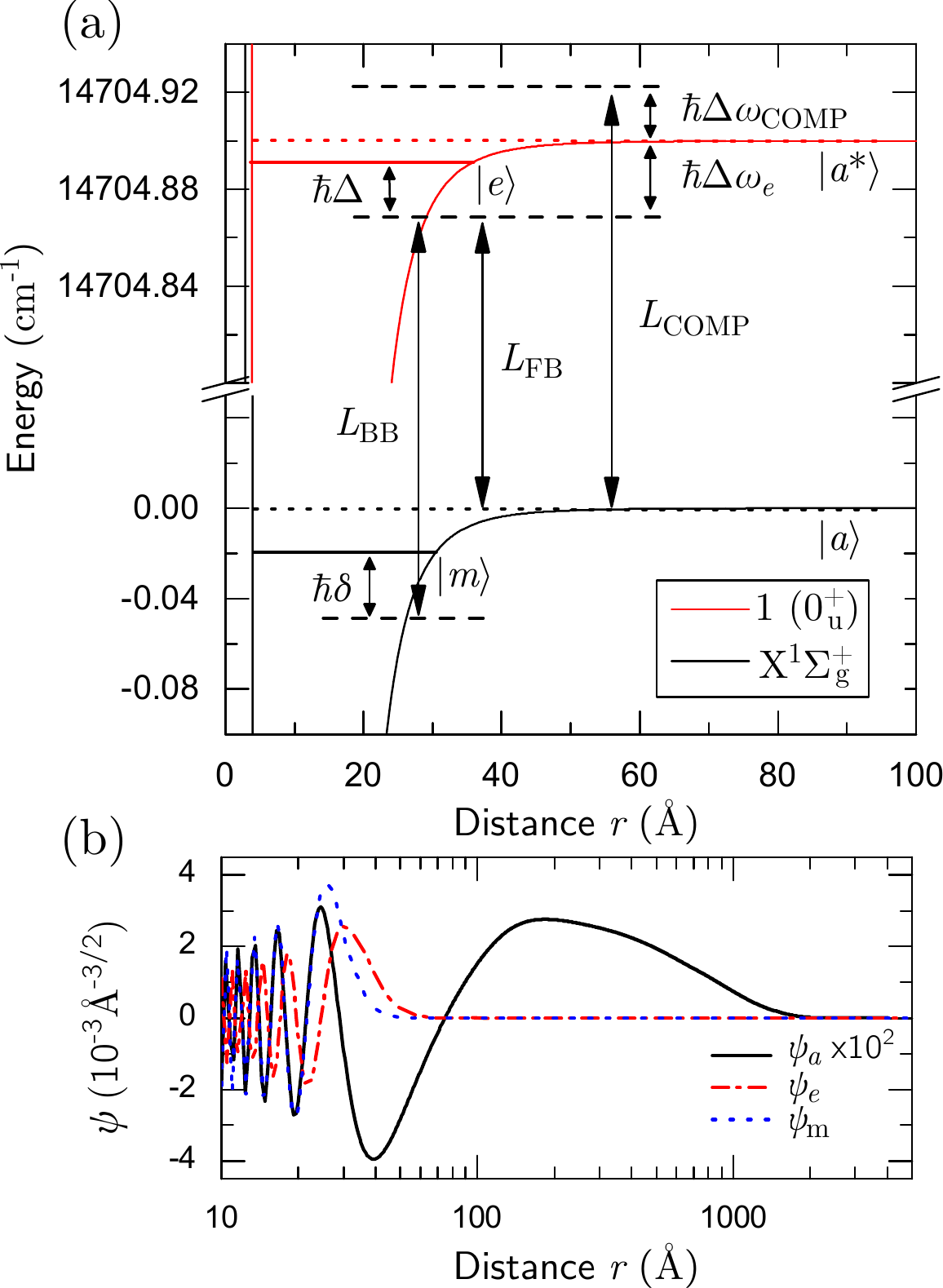}
\caption{\label{fig:IntroPicture} (color online) (a) $^{84}\mathrm{Sr}_{2}$ molecular potential for the electronic ground state $X^{1}\Sigma^{+}_{g}$ ($J=0$) and the optically exited state $1(0^{+}_{u})$ ($J=1$). The energy is referenced to the ground-state asymptote. The $\Lambda$ scheme $( \vert a \rangle , \vert e \rangle, \vert m \rangle )$ used for STIRAP is indicated along with the coupling laser fields $L_{\mathrm{FB}}$ and $L_{\mathrm{BB}}$ and their one- and two-color detunings $\Delta$ and $\delta$. The laser field $L_{\mathrm{COMP}}$ with detuning $\Delta \omega_{\mathrm{COMP}}$ from the $\vert a \rangle - \vert a^* \rangle$ atomic transition is added for light-shift compensation in Sec.~\ref{subsec:Light-shift compensation}. In the situation shown $\Delta$, $\delta$, $\Delta \omega_{\mathrm{COMP}}>0$ and $\Delta \omega_e<0$. (b) Probability amplitude for the nuclear wavefunction of state $\vert a \rangle$, $\vert e \rangle$ and $\vert m \rangle$. The effect of an external harmonic confinement is taken into account to determine $\psi_{a}$, see Sec.~\ref{subsec:MI STIRAP}.  }
\end{figure}

To optically transfer a pair of atoms in state $\vert a \rangle$ into the molecular state $\vert m \rangle$, we use a $\Lambda$ scheme, including an optically excited molecular state $\vert e \rangle$, see Fig.~\ref{fig:IntroPicture}. States $\vert a \rangle$ and $\vert e \rangle$ are coupled with Rabi frequency $\Omega_{\mathrm{FB}}$ by the free-bound laser $L_{\mathrm{FB}}$ and $\vert e \rangle$ is coupled to $\vert m \rangle$ with Rabi frequency $\Omega_{\mathrm{BB}}$ by the bound-bound laser $L_{\mathrm{BB}}$.

The conceptually simplest method for coherent molecule association are two consecutive $\pi$ pulses, the first between $\vert a \rangle$ and $\vert e \rangle$ and the second between $\vert e \rangle$ and $\vert m \rangle$. To provide efficient transfer, this scheme needs to be executed much faster than the lifetime of the excited state $\vert e \rangle$, $\tau_e$, which requires high Rabi frequencies ($\Omega_{\mathrm{FB,BB}} \gg \gamma_e=1/\tau_e$). Since the Franck-Condon factor (FCF) of the free-bound transition is small, satisfying the condition $\Omega_{\mathrm{FB}} \gg \gamma_e$ is experimentally very challenging.

STIRAP overcomes this limitation by minimizing losses from $\vert e \rangle$ and provides coherent population transfer even when the condition $\Omega_{\mathrm{FB,BB}} \gg \gamma_e$ is not satisfied. To simplify the discussion we first introduce STIRAP without loss from the excited state ($\gamma_e = 0$) and with lasers $L_{\mathrm{FB}}$ and $L_{\mathrm{BB}}$ on resonance with the transitions $\vert a \rangle - \vert e \rangle$ and $\vert m \rangle - \vert e \rangle$, respectively. In this system a dark state exists, namely an eigenstate orthogonal to $\vert e \rangle$. If only one of the two lasers ($L_{\mathrm{FB}}$ or $L_{\mathrm{BB}}$) is on, then the dark state coincides with one of the eigenstates in the absence of light ($\vert m \rangle$ or $\vert a \rangle$, respectively). When both lasers are on, the dark state is $\mathrm{\vert dark \rangle} = \mathrm{\cos(\theta)\, \vert a \rangle - \sin(\theta) \, \vert m \rangle} $ with $\theta= \arctan\mathrm{(\Omega_{FB}/\Omega_{BB})}$. If the tuning knob $\theta$ is varied in time from $0$ to $\pi /2$, then the dark state is moved from $\vert a \rangle$ to $\vert m \rangle$. Experimentally this is realized by changing the intensities $I_{\mathrm{FB,BB}}$ of $L_{\mathrm{FB,BB}}$, first switching on $L_{\mathrm{BB}}$, then increasing $I_{\mathrm{FB}}$ while ramping off $I_{\mathrm{BB}}$, and finally switching off $L_{\mathrm{FB}}$. Since an energy gap exists between the dark state and the other two instantaneous eigenstates of the system, the adiabatic theorem of quantum mechanics applies \cite{Born1928AdiabTheo, Kato1950AdiabTheo}, provided that the change in the Hamiltonian is slow enough compared to the energy gap. This means that the population is kept in the dark state, providing unit transfer efficiency. State $\vert e \rangle$ is only used to induce couplings but never significantly populated. As a consequence this scheme works even in the presence of dissipation ($\gamma_e \neq 0$) as will be explained in Sec.~\ref{sec:Model and parameters}.

In order to detect that atoms have been associated to molecules, we exploit the fact that STIRAP is reversible. After the association STIRAP (aSTIRAP) described above, we push remaining free atoms away with a pulse of resonant light. Molecules are barely affected by this light pulse because they do not have any strong optical transition close in frequency to the atomic transition used. To detect the molecules, we dissociate them by a time-mirrored aSTIRAP, which we call dissociation STIRAP (dSTIRAP), and detect the resulting atoms by absorption imaging. The sequence ``aSTIRAP --- push pulse --- dSTIRAP" constitutes a STIRAP cycle and will be used to prepare samples with homogeneous conditions for molecule association and to measure the STIRAP efficiency, see Sec.~\ref{subsec:MI STIRAP}.

We exploit a deep optical lattice to create a MI of doubly occupied sites with high on-site peak density. The reason for this is twofold. Firstly, the optical lattice increases the free-bound Rabi frequency, see Sec.~\ref{subsubsec: Rabi frequencies}, leading to an increase in the STIRAP efficiency. Secondly, the lattice increases the molecule lifetime by suppressing collisional losses, see Sec.~\ref{subsubsec: Molecule lifetime}, leading both to an increase in the STIRAP efficiency and to a longer lifetime of the resulting molecular sample.

The $\Lambda$ scheme used here is the same as the one employed in our previous work \cite{Stellmer2012Sr2Mol}. The relevant potential energy curves of $^{84}\mathrm{Sr}_{2}$ are shown in Fig.~\ref{fig:IntroPicture}. The electronic ground state is a $X^{1}\Sigma^{+}_{g}$ ($J=0$) state, asymptotically correlating to two ground-state Sr atoms (${^1S_0}$), and the excited state is a $1(0^{+}_{u})$ ($J=1$) state, correlating to one Sr atom in the ground state and one in the optically excited state ${^3P_1}$. The initial state $\vert a \rangle$ consists of two atoms in the lowest vibrational level of an optical lattice well. At short range, their relative-motion wavefunction is proportional to a scattering state above the dissociation threshold of $X^{1}\Sigma^{+}_{g}$ ($J=0$). Our target state $\vert m \rangle$ is the second to last vibrational level ($\nu=-2$) supported by this potential, with binding energy $E_{m}= h \times \unit[644.7372(2)]{MHz}$, where $h$ is the Planck constant. As intermediate state $\vert e \rangle$ we chose the $\nu=-3$ vibrational level of the $1(0^{+}_{u})$ potential, with binding energy $E_{e}= h \times \unit[228.38(1)]{MHz}$. The STIRAP lasers are appropriately detuned from the ${^1S_0} - {^3P_1}$ transition of Sr at $\unit[689]{nm}$, which has a linewidth of $\Gamma_{^3{\rm P}_1} = 2 \pi \times \unit[7.4]{kHz}$. For the push beam we use the ${^1S_0} - {^1P_1}$ transition at $\unit[461]{nm}$, which has a linewidth of $\Gamma_{^1{\rm P}_1} = 2 \pi \times \unit[30.5]{MHz}$.

\section{Theory}
\label{sec:Model and parameters}

In this section we will introduce a model for molecule association via STIRAP. We will discuss the conditions on experimental parameters under which STIRAP is efficient. An important insight will be that finite two-color detuning $\delta$ can lead to a significant reduction in the dark state lifetime and therefore efficiency. We give an overview of the sources for this detuning in our experiment and ways to overcome it.

\subsection{Model}
\label{subsec:Model}

A model for STIRAP starting from two atoms in the ground state of an optical lattice is given by the time-dependent Schr\"odinger equation applied to the three relevant states, which in the rotating wave approximation takes the form \cite{Vitanov1997StirapPopTransferTheory,Winkler2007CoherentFeshbachMol}

\begin{equation}
\label{MI Model}
i \frac{d}{dt}
\begin{pmatrix}
  a\\
  e\\
  m
 \end{pmatrix}=-\frac{1}{2}
 \begin{pmatrix}
  i \gamma_{a} & \Omega_{\mathrm{FB}} & 0\\
  \Omega_{\mathrm{FB}} & i\gamma_{e}-2\Delta  & \Omega_{\mathrm{BB}}\\
  0 & \Omega_{\mathrm{BB}} & i \gamma_{m}-2\delta
 \end{pmatrix}
 \begin{pmatrix}
  a\\
  e\\
  m
 \end{pmatrix}
 .
\end{equation}

\noindent The amplitudes $a(t)$, $e(t)$, and $m(t)$ correspond to the states $\vert a \rangle$, $\vert e \rangle$, and $\vert m \rangle$, respectively. Losses from these states are described by $\gamma_{a}$, $\gamma_{e}$ and $\gamma_{m}$, which are responsible for the coupling of the three-level system to the environment.  When loss rates are set to zero, the normalization condition is $\vert a \vert ^2 + \vert e \vert^2+ \vert m \vert^2=1$. Finally, $\Delta$ and $\delta$ are the one- and two-color detunings, see Fig.~\ref{fig:IntroPicture}. In the remainder of this section we will study the STIRAP efficiency under the following assumptions. STIRAP is executed during the time interval $[0,T]$, outside of which the two Rabi frequencies $\Omega_{\mathrm{FB,BB}}$ are zero. For times $t\in [0,T]$ the Rabi frequencies are described by (co)sine pulses of identical amplitude $\Omega_m$, $\Omega_{\mathrm{BB}}=\Omega_m \cos(\theta)$ and $\Omega_{\mathrm{FB}}=\Omega_m \sin(\theta)$, with $\theta=\pi t/2T$. The scattering rate $\gamma_{e}$ is the dominant loss term in the system, i.e. $\gamma_{e}\gg \gamma_{a,m} $. The two-color detuning depends on time only through $\theta$, i.e. $\delta = \delta (\theta)=\delta_m f(\theta)$, where $\delta_m$ is the (signed) extremum of $\delta (\theta)$ with highest amplitude, and $f(\theta)$ describes the time variation caused by the time-dependent light shifts discussed in Sec.~\ref{subsec:Improved STIRAP}. Finally, unless stated otherwise, we will assume $\Delta=0$.

\subsection{Parameter constraints}
\label{subsec:Parameter constraints}

We now analyse the constraints on the (experimentally controllable) detunings and Rabi frequencies under which STIRAP associates molecules with near-unit efficiency. We will first consider the resonant case (zero detunings) and derive the parameter constraints analytically, following Ref. \cite{Vitanov1997StirapPopTransferTheory}. We will then study the effects of deviations from two-color resonance, where an approximate solution and new parameter constraints will be presented. Finally, the effects of one-color and two-color detunings will be compared.

We analyse the problem in the experimentally relevant case of strong dissipation $\Omega_m\lesssim \gamma_{e}$, in which STIRAP has the potential to outperform two appropriate consecutive $\pi$-pulses. Here dissipation leads to a strong non-hermitian contribution in the hamiltonian \cite{Nenciu1992AdiabTheoNonAdjoint}. By adiabatically eliminating the variable $e(t)$ as in Ref. \cite{Vitanov1997StirapPopTransferTheory}, we derive the effective Hamiltonian $\widetilde{H}_{\rm eff}$ for the subspace $\left\{ \vert a \rangle,\vert m \rangle \right\}$,

\begin{equation}
\label{MI Eff Model}
\widetilde{H}_{\rm eff}=-i\frac{\tilde{\gamma}}{2}
 \begin{pmatrix}
  \sin(\theta)^2 & \sin(\theta)\cos(\theta) \\
  \sin(\theta)\cos(\theta) & \cos(\theta)^2+2iAf(\theta)\\
 \end{pmatrix}
 ,
\end{equation}

\noindent where $\tilde{\gamma}=\Omega_m^2 / \gamma_{e}$ and $A=\delta_m/ \tilde{\gamma}$.

We first examine the case $\delta = \Delta =0$ and we will show that adiabatic evolution ensures near-unit transfer efficiency \cite{Vitanov1997StirapPopTransferTheory}. Despite the strong dissipation, this system supports the same dark state as before, the eigenstate $ \vert \mathrm{dark} \rangle = \cos(\theta)\, \vert a \rangle - \sin(\theta) \, \vert m \rangle $. The orthogonal lossy state is the bright state $\vert \mathrm{bright} \rangle$. In order to gain some insight, we work in the adiabatic representation, i.e. in the instantaneous eigenbasis $\left\{ \vert \mathrm{dark} \rangle, \vert \mathrm{bright} \rangle \right\}$ of the Hamiltonian (\ref{MI Eff Model}). A wavevector $\vert \Psi,t \rangle$ in the basis $\left\{ \vert a \rangle,\vert m \rangle \right\}$ is written $\vert \Psi^{'},t \rangle= \widetilde{U}(t) \vert \Psi,t \rangle$ in this new basis, with the appropriate unitary transformation $\widetilde{U}$. The time-dependent Schr\"odinger equation reads $ i \frac{d}{dt}\vert \Psi^{'} \rangle = \widetilde{U} \widetilde{H}_{\rm eff} \widetilde{U}^{-1} \vert \Psi^{'} \rangle + i \left( \frac{d}{dt} \widetilde{U} \right) \widetilde{U}^{-1} \vert \Psi^{'} \rangle= \widetilde{H}^{'} \vert \Psi^{'} \rangle$, where $\widetilde{U} \widetilde{H}_{\rm eff} \widetilde{U}^{-1}$ is diagonal and $\left( \frac{d}{dt} \widetilde{U} \right) \widetilde{U}^{-1}$ gives rise to non-adiabatic couplings. If we take $s=t/T \in \left[ 0,1 \right]$ as our independent variable, the Schr\"odinger equation becomes $ i \frac{d}{ds}\vert \Psi^{'} \rangle =T \widetilde{H}^{'} \vert \Psi^{'} \rangle$. In the case of zero detunings, $T \widetilde{H}^{'}$ is independent of $s$ and the equation is exactly solvable. The solution for the dark-state amplitude $d(s)$ is $d(s)=B_{-} e^{-i \lambda_{-}s}+B_{+} e^{-i \lambda_{+}s}$, where $\lambda_{-,+}$ are respectively the most and least dissipative eigenvalues of $T \widetilde{H}^{'}$, and $B_{-,+}$ are constants. In the adiabatic regime, characterized by $\alpha = \tilde{\gamma} T \gg 1$, we have $\mathrm{Im}\left( \lambda_{-} \right)\simeq -\alpha/2$ and $\mathrm{Im}\left( \lambda_{+} \right)\simeq -\pi^2 / 2 \alpha$. The transfer efficiency $\eta$ is then given by $\eta=\vert d(s=1)\vert^2\simeq \vert B_{+}\vert^2 e^{2 \lambda_{+}}\simeq e^{-\pi^2 /\alpha}$. Thus, STIRAP with near-unit efficiency requires $\alpha\gg \pi^2$, which can be interpreted as the adiabaticity condition.

Next we examine the case of non-zero two-color detuning, i.e. $\delta \neq 0$, and we show that adiabatic evolution does not guarantee high transfer efficiency. Going from zero to finite $\delta$, the initially dark state is mixed with $\vert e \rangle$, resulting in a finite lifetime. For simplicity we all the same continue to use the label ``dark state" for this state and, similarly, keep using ``bright state" for the other eigenstate. We apply the same method as used previously, with $\widetilde{U}$ being this time the appropriate non-unitary transformation. The time evolution of the system is determined by $T \widetilde{H}^{'}$, which now depends on $s$ and the solution is not trivial. However, since we are only interested in small $\delta$ and the populations at the end of the pulse, we replace $T \widetilde{H}^{'}$ with its time average. The imaginary parts of the most and least dissipative eigenvalues are, to leading order in $A^n/ \alpha^m$, $\mathrm{Im}\left( \lambda_{-} \right)\simeq - \alpha/2$ and $\mathrm{Im}\left( \lambda_{+} \right)=-1/2(\pi^2 /\alpha + C A^2 \alpha)$, where $C=\frac{8}{\pi} \int_0^{\pi/2}f(\theta)^2 \sin(\theta)^2 \cos(\theta)^2 d\theta$. Thus, the transfer efficiency is approximately given for $\alpha \gg 1$, $ \vert A \vert \ll 1$ by

\begin{equation}
\label{STIRAP Eff}
\eta \simeq e^{-(\pi^2 /\alpha + C A^2 \alpha)}
,
\end{equation}

\noindent which is only a function of the parameters $\alpha$ and $A$. We identify two regimes depending on the value of  $\vert A \vert \, \alpha = \vert \delta_m \vert T$. For $  \vert \delta_m \vert T \ll \pi / \sqrt{C} $, the transfer efficiency is limited by non-adiabatic transitions of the dark state to the bright state. For $ \vert \delta_m \vert T \gg \pi / \sqrt{C} $, the main limitation arises from the now finite lifetime of the dark state. For a non-zero detuning $\delta$, the transfer efficiency features a maximum when varying pulse time $T$, which is determined only by the parameter $\vert A\vert=\vert\delta_m\vert/ \tilde{\gamma}$, and the maximum efficiency is approximately $\eta_\mathrm{max}=e^{-2 \pi \sqrt{C} \vert A \vert}$.

We now address the case of non-zero one-color detuning $\Delta \neq 0$, but zero two-color detuning $\delta=0$, and we show that, as for $\delta =\Delta =0$, adiabatic evolution ensures near-unit transfer efficiency. In this case, the Hamiltonian (\ref{MI Model}) supports a dark state, and by following the method used for $\delta=\Delta=0$, it can be shown that $T\widetilde{H}^{'}$ is independent of $s$. The eigenvalues of $T \widetilde{H}^{'}$ are, to leading order in $1/\alpha$, given by $\mathrm{Im}\left( \lambda_{-} \right)\simeq -\alpha/2 \left( 1+4 (\Delta / \gamma_e)^2 \right)$ and $\mathrm{Im}\left( \lambda_{+} \right)\simeq -\pi^2 /2 \alpha$. This implies that, even for $\Delta\neq 0$, we can have $\vert \mathrm{Im}\left( \lambda_{+} \right)\vert \ll 1$ and thus $\eta\simeq 1$ for large enough values of $\alpha$. To compare the effects of one-color and two-color detunings, we analyze the conditions to ensure high transfer efficiency at large but fixed $\alpha$. For a constant one-color detuning we derive $\Delta \ll \gamma_e \alpha/ 2 \pi$, while for a constant two-color detuning we have $\vert \delta_m \vert \ll \tilde{\gamma} \sqrt{2/\alpha} \lesssim \gamma_e \sqrt{2/\alpha}$, which is a much stronger constraint. The STIRAP efficiency is therefore much more sensitive to the two-color detuning than to the one-color detuning, which is also expected from energy conservation.

It is noteworthy to point out that population transfer between  $\vert a \rangle$ and $\vert m \rangle$ can be obtained by exploiting the coherent dark state superposition in a \textit{diabatic} evolution, i.e. projection, as opposed to STIRAP. However the efficiency in this case is bound to be at most $\eta_{\mathrm{proj}}= \unit[25]{\%}$, which is the value for equal free-bound and bound-bound Rabi frequencies, as obtained from $\eta_{\mathrm{proj}}= \vert \langle \mathrm{dark} \vert a \rangle \vert ^2 \times \vert \langle m \vert \mathrm{dark} \rangle \vert ^2= \left( 1- \sin(\theta)^2 \right) \sin(\theta)^2 \leq 1/4$. As a consequence surpassing the maximum possible efficiency of $\eta_{\mathrm{proj}}$ can be regarded as a requirement for STIRAP to be relevant.

\subsection{Improving STIRAP efficiency}
\label{subsec:Improved STIRAP}

Finally, we discuss how we can improve the molecule production efficiency based on this theoretical description. We will focus on the limitations arising from light shifts induced on the states of the $\Lambda$ scheme that contribute to $\delta$ and $\Delta$. We distinguish between $static$ and $dynamic$ light shifts, where static and dynamic refer to their behavior during the STIRAP sequence. The lasers $L_{\mathrm{FB}}$ and $L_{\mathrm{BB}}$, whose intensities vary during STIRAP, induce time- and space-dependent shifts $\Delta_{\mathrm{FB}}$ and $\Delta_{\mathrm{BB}}$ on the free-bound transition. In a similar way, these lasers also induce shifts $\delta_{\mathrm{FB}}$ and $\delta_{\mathrm{BB}}$ on the two-photon transition. These shifts correspond to changes $-\hbar \delta_{\mathrm{FB, BB}}$ of the binding energy of molecules in state $\vert m \rangle$, which is the energy difference between $\vert a \rangle$ and $\vert m \rangle$. The optical lattice, which provides the external confinement, induces only the static, space-dependent light shifts $\Delta_{\mathrm{Lattice}}$ and $\delta_{\mathrm{Lattice}}$, respectively, onto the free-bound transition and the two-photon transition. Thus, we decompose $\Delta$ and $\delta$ into $\Delta=\Delta_{\mathrm{Lattice}} + \Delta_{\mathrm{FB}} + \Delta_{\mathrm{BB}} + \Delta_0$ and $\delta=\delta_{\mathrm{Lattice}} + \delta_{\mathrm{FB}} + \delta_{\mathrm{BB}} + \delta_0$. The offsets $\Delta_0$ and $\delta_0$ can be freely adjusted by tuning $L_{\mathrm{FB,BB}}$ and depend neither on space nor on time.

These light shifts can influence the molecule production efficiency $\eta$. In our system the effect of $\Delta$ on $\eta$ is negligible, see Sec.~\ref{subsubsec: Determination of gammae}, justifying the approximation $\Delta=0$. Since we also fulfill $\gamma_{a,m}\ll \gamma_{e}$ (see Sec.~\ref{subsubsec: Light shifts and scattering from PA light} and \ref{subsubsec: Molecule lifetime}), we can use Eqn.~\ref{STIRAP Eff}, which shows that the maximum transfer efficiency for optimal pulse time depends only on $\vert A \vert=\vert\delta_m\vert/ \tilde{\gamma}$. The efficiency is highest for small $\vert A \vert$, which means we have to achieve low $\vert\delta_m\vert$. The contribution $\delta_{\mathrm{Lattice}}$ to $\delta_m$ could be reduced by increasing the lattice beam diameter and adjusting $\delta_0$ to compensate for the average remaining light shift across the sample. However in this work we reduce the inhomogeneity by simply removing atoms on sites with large light shift using a STIRAP cycle, see Sec.~\ref{subsec:MI STIRAP} and Sec.~\ref{subsubsec: STIRAP optimization}. The contribution $\delta_{\mathrm{FB}}$ is proportional to $I_{\mathrm{FB}}\propto \sin(\theta)^2$. Lowering $A=\delta_m/ \tilde{\gamma}$ by lowering $I_{\mathrm{FB}}$ is not possible since also $\tilde{\gamma}\propto I_{\mathrm{FB}}$. A straightforward way to half the effect of this contribution is to adjust $\delta_0$ such that $\delta = \delta_m ( \sin(\theta)^2 - 1/2 )$. To go further and essentially cancel $\delta_{\mathrm{FB}}$ we demonstrate in Sec.~\ref{subsec:Light-shift compensation} the use of an additional laser beam ($L_{\mathrm{COMP}}$) that produces the exact opposite light shift of $L_{\mathrm{FB}}$. The last light shift, $\delta_{\mathrm{BB}}$, can be neglected in our case because $L_{\mathrm{BB}}$ has low intensity.

\section{Experimental setup and creation of the Mott insulator}
\label{sec:Ingredients}

We will now describe the experimental setup and the procedure used to prepare the atomic sample that is the starting point for molecule creation, and describe the laser system that generates the PA light.

The experimental apparatus is our Sr quantum gas machine, which is described in depth in \cite{Stellmer2013QDegSr}. The experimental sequence leading to a Bose-Einstein condensate (BEC) of $^{84}\mathrm{Sr}$ starts with a magneto-optical trap operating on the ${^1S_0} - {^1P_1}$ transition, capturing atoms coming from a Zeeman slower. We further cool the atomic cloud by exploiting the intercombination transition ${^1S_0} - {^3P_1}$ at $\unit[689]{nm}$, reaching a temperature of $\unit[1.0]{\mu K}$. We then load the atoms into a crossed-beam dipole trap (DT), consisting of one horizontal beam with vertical and horizontal waists of $\unit[20]{\mu m}$ and $\unit[330]{\mu m}$, respectively, and one near-vertical beam with a waist of $\unit[78]{\mu m}$. After evaporative cooling we obtain a BEC of about $3.5 \times 10^5$ atoms, with a peak density of $\unit[6.0(5)\times 10^{13}]{cm^{-3}}$. The trapping frequencies are $\omega_{x}= \unit[2\pi \times 16]{Hz}$, $\omega_{y}= \unit[2\pi \times 11]{Hz}$, $\omega_{z}= \unit[2\pi \times 95]{Hz}$, where the $y$-axis points along the horizontal DT beam and the $z$-axis is vertical. The Thomas-Fermi radii are $R_{x}= \unit[23]{\mu m}$, $R_{y} = \unit[35]{\mu m}$ and $R_{z}= \unit[4.1]{\mu m}$, and the chemical potential is $\mu = \unit[30(2)]{nK}$. These parameters are chosen to maximize the number of doubly-occupied sites in the optical lattice.

The Mott insulator (MI) is realized by adiabatically loading the BEC into a 3D optical lattice, whose standing-wave interference pattern is obtained from three orthogonal, retro-reflected laser beams, with waists of $\unit[78]{\mu m}$, $\unit[206]{\mu m}$ and $\unit[210]{\mu m}$, derived from a single-mode laser with $\lambda_{\mathrm{Lattice}}=\unit[1064]{nm}$ wavelength (Innolight Mephisto MOPA). Orthogonal polarizations and frequency offsets of about $\unit[10]{MHz}$ between the three laser beams ensure that the optical potential experienced by the atoms can be well described by the sum of three independent 1D lattices. The lattice depth is calibrated through Kapitza-Dirac diffraction \cite{Ovchinnikov1999DiffractionBECLattice}.

The BEC is adiabatically loaded into the optical lattice by increasing the lattice potential in three consecutive exponential ramps. A first ramp of $\unit[500]{ms}$ duration is used to reach a depth of about $\unit[5]{E_r}$ in each lattice direction, where $\mathrm{E_r} = h \times \unit[2.1]{kHz}$ is the recoil energy. During the second ramp, which takes $\unit[150]{ms}$ and reaches $\unit[20]{E_r}$, the system undergoes the superfluid to MI phase transition. We then switch off the crossed-beam DT. The final ramp increases the trap depth to $\unit[200]{E_r}$ in $\unit[5]{ms}$ in order to reduce the size of the lattice site ground-state wavefunction, which in turn increases the free-bound Rabi frequency, see Sec.~\ref{subsubsec: Rabi frequencies}.

Since only atoms in doubly occupied sites will contribute to molecule formation, we maximize that number by varying initial atom number and DT confinement. To determine the number of doubly-occupied sites, we exploit three-body recombination and photoassociation. After loading the lattice, three-body recombination empties triply-occupied sites on a $1/e$ timescale of $\unit[104(15)]{ms}$. After $\unit[200]{ms}$ of wait time, we also empty the doubly-occupied sites by PA using $L_{\mathrm{FB}}$. The difference in atom number before and after the PA pulse is twice the number of doubly-occupied sites. Measuring the atom number also directly after the loading sequence we derive the occupation fractions $\unit[47(9)]{\%}$, $\unit[30(5)]{\%}$ and $\unit[23(4)]{\%}$ for the occupation numbers $n=1$, $n=2$ and $n=3$, respectively, where we assumed that the number of sites with more than three atoms is negligible. To prepare samples for molecule association by STIRAP, we wait 200\,ms after loading the lattice, thereby removing most sites with three or more atoms.

\begin{figure}[tb]
\includegraphics[width=\columnwidth]{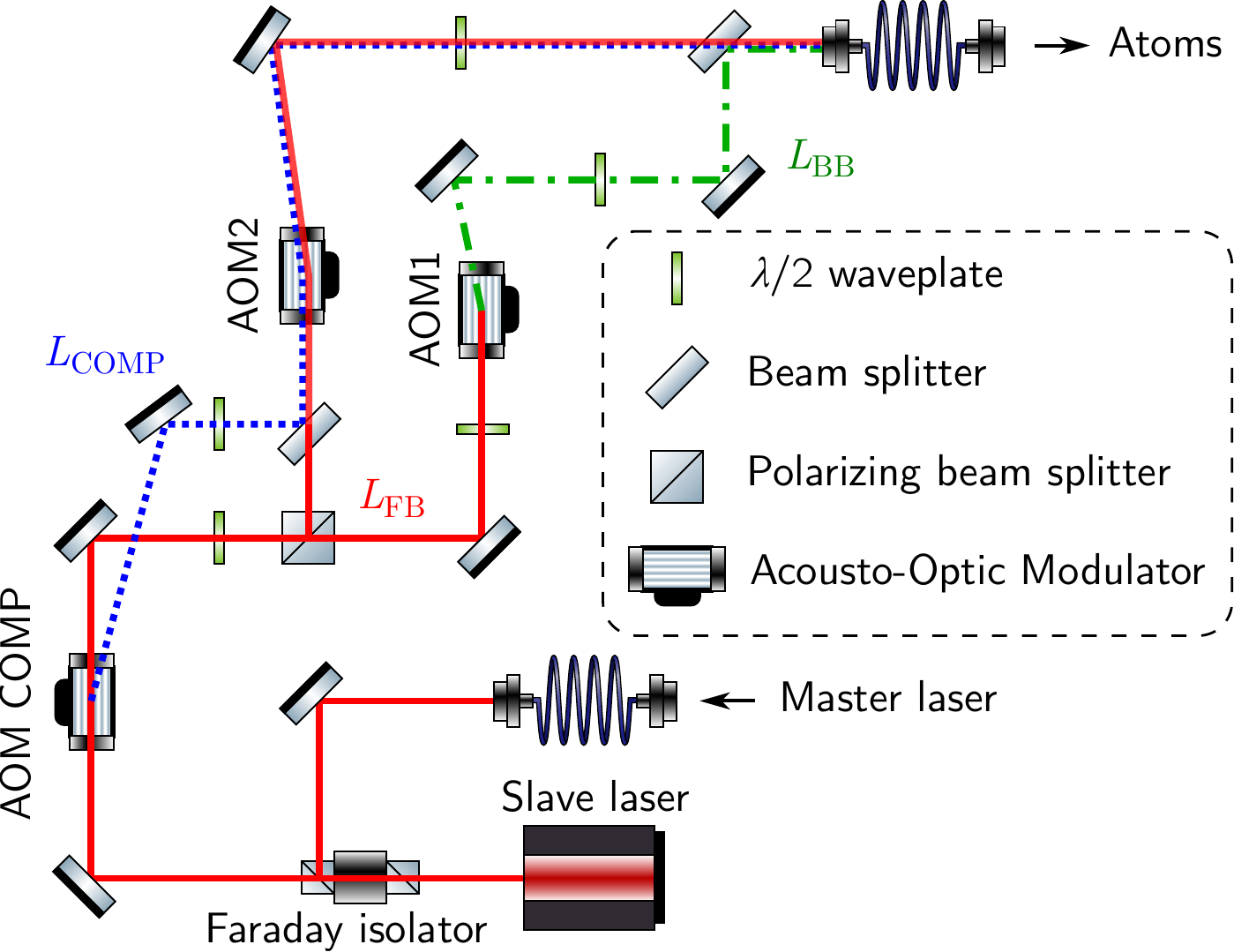}
\caption{\label{fig:LevelSchemeCompensation} (color online) Optical setup used to produce photoassociation light and to compensate the light shift $\delta_{\mathrm{FB}}$ in both space and time.}
\end{figure}

We now describe the laser setup used to illuminate the atoms with PA light, see Fig.~\ref{fig:LevelSchemeCompensation}, which also generates the light-shift compensation beam $L_{\mathrm{COMP}}$, see Sec.~\ref{subsubsec:Compensation beam}. The two laser fields $L_{\mathrm{FB}}$ and $L_{\mathrm{BB}}$ are derived, by free-space splitting and recombination, from a single injection-locked slave laser seeded by a master oscillator with linewidth of less than $\unit[2\pi \times 3]{kHz}$. The frequencies of the laser fields are tuned by acousto-optical modulators (AOMs) and the beams are recombined into the same single-mode fiber with the same polarization, so that the main differences on the atomic cloud are their frequency and intensity. This setup ensures a good coherence between the two laser fields, which must match the Raman condition, which means that the frequency difference between the lasers must be equal to the binding energy of state $\vert m \rangle$ divided by $h$. The quality of the beat note of $L_{\mathrm{FB}}$ with $L_{\mathrm{BB}}$ is essentially set by the electronics controlling the AOMs. In the spectrum of the beat note, recorded on a photodiode and analyzed with a bandwidth of $\unit[2\pi \times 3]{Hz}$, the beat signal rises by $\sim \unit[60]{dB}$ above the background and has a width of $\unit[2\pi \times 60]{Hz}$. This narrow width can be neglected on the time scale of the experiment (hundreds of $\unit{\mu s}$) and doesn't need to be taken into account in our theoretical model of STIRAP. Frequency fluctuations of the master laser do not change the two-color detuning $\delta$. They only change the one-color detuning $\Delta$ to which the STIRAP efficiency has low sensitivity, see Sec.~\ref{subsec:Parameter constraints}. Finally, all laser fields used for PA are contained in one beam, which is sent onto the atomic cloud horizontally, under an angle of $\unit[30]{^{\circ}}$ from the $y$-axis. The waist of the beam at the location of the atoms is $\unit[113(2)]{\mu m}$. The polarization is linear and parallel to a vertically oriented guiding magnetic field of $\unit[5.30(5)]{G}$, which means that only $\pi$ transitions can be addressed. The magnetic field splits the Zeeman levels of the state $\vert e \rangle$ by $ \unit[2\pi \times 1.65(1)]{MHz} \gg \gamma_{e}$.

\section{Molecule creation}
\label{sec:Molecule creation}

We will now discuss molecule creation via STIRAP. First, we will characterize the parameters that can be measured before attempting STIRAP (PA Rabi frequencies, dynamic light shifts from PA light, and static light shifts from the lattice), see Tab.~\ref{table:ParametersMI}. We will then apply STIRAP on our sample and identify the finite lifetime of the dark state arising from the light shift $\delta_{\mathrm{FB}}$ as the main limitation to the STIRAP efficiency. We will show how to overcome this limitation by minimizing $\delta_{\mathrm{FB}}$ using a compensation beam and we will examine the effects of this compensation scheme on STIRAP. Finally we will characterize the spectral properties of the initial atomic sample and the lifetime of the molecular sample.

\subsection{Parameter characterization}
\label{subsec:MI Parameters}

\subsubsection{Rabi frequencies}
\label{subsubsec: Rabi frequencies}

We measure the bound-bound Rabi frequency $\Omega_{\mathrm{BB}}$ through loss spectroscopy by probing the Autler-Townes splitting induced by $\Omega_{\mathrm{BB}}$ with the free-bound laser \cite{Autler1955AutlerTownesSplitting}. We derive from our measurements $\Omega_{\mathrm{BB}}= \unit[2\pi \times 234(5)]{kHz / \sqrt{W/cm^{2}} }$.

\begin{figure}[tb]
\includegraphics[width=\columnwidth]{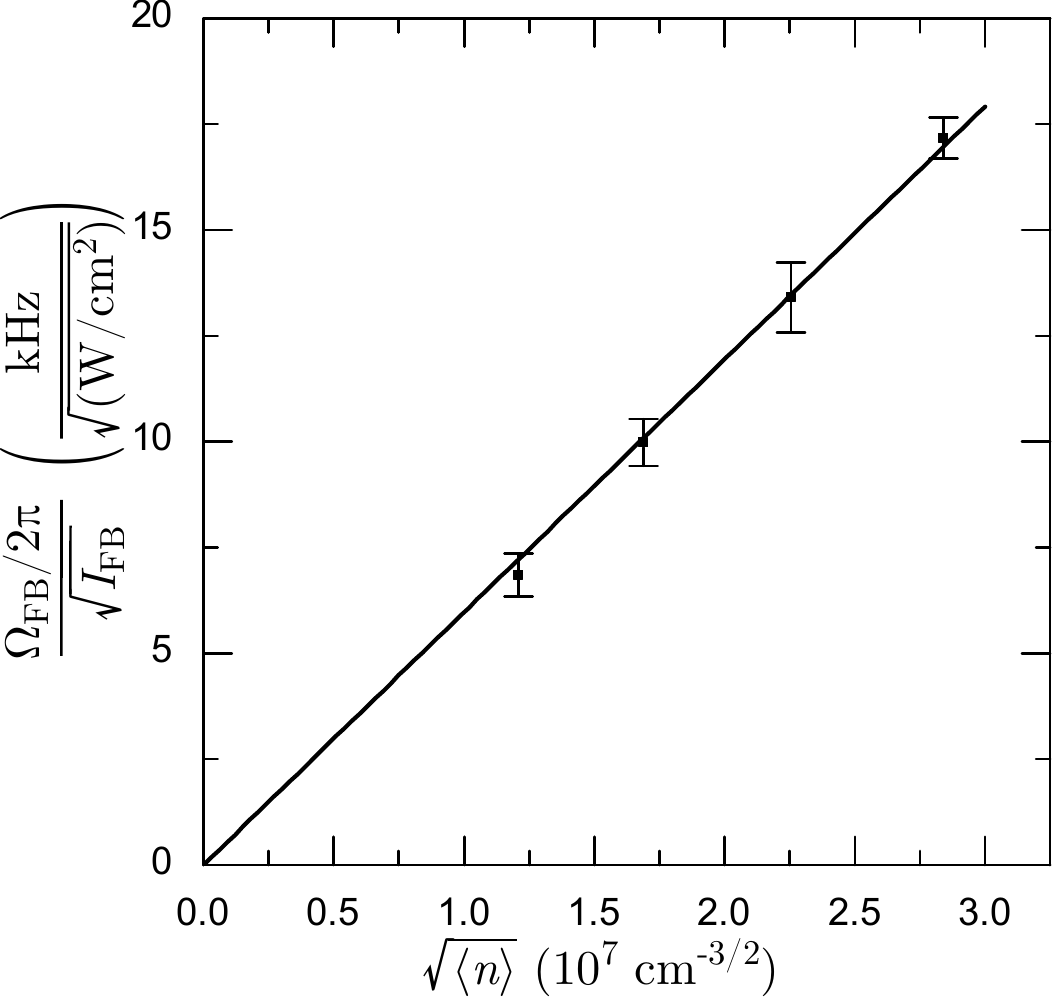}
\caption{\label{fig:FBOmegaVSExtConf} Free-bound Rabi frequency $\Omega_{\mathrm{FB}}$ as function of $\sqrt{\langle n \rangle}$, where $\langle n \rangle$ is the average on-site density of a single atom. The line is a linear fit to the data, justified in the text.}
\end{figure}

To measure the free-bound Rabi frequency $\Omega_{\mathrm{FB}}$ we shine $L_{\mathrm{FB}}$ on the MI and detect the resulting decay of atom number as a function of time. The observed decays are well described by exponentials with time constants $\tau$. We measure $\tau$ for several lattice depths and intensities of $L_\mathrm{FB}$. The variation of the atom number $N_a \propto  \vert a\vert^2$ is determined by $\dot{a}=-\frac{\Omega_{\mathrm{FB}}^2}{2 \Gamma_e}\,a$, which is obtained by adiabatic elimination of variable $e$ in model (\ref{MI Model}) and by using the natural linewidth $\Gamma_e$ of the free-bound transition as $\gamma_e$. Thus the measured decay time constant $\tau$ can be related to the Rabi frequency by $\Omega_{\mathrm{FB}}=\sqrt{\Gamma_e / \tau}$. The measurement of $\Gamma_e$ is explained in Sec.~\ref{subsubsec: Inhomogeneous light shifts from lattice light}. Figure~\ref{fig:FBOmegaVSExtConf} shows the free-bound Rabi frequency $\Omega_{\mathrm{FB}}$ as a function of $\sqrt{\langle n \rangle}$, where $n$ is the on-site density of a single atom and $\langle \cdot \rangle$ is the spatial average over one site. We observe that $\Omega_{\mathrm{FB}}$ is proportional to $\sqrt{\langle n \rangle}$, and we derive  $\Omega_{\mathrm{FB}}= \unit[2 \pi \times 6.0(1) \, 10^{-7}]{kHz \, cm^{3/2} / \sqrt{W/cm^{2}} }$.

We now present an argument that explains the observed linear relationship between $\Omega_{\mathrm{FB}}$ and $\sqrt{\langle n \rangle}$ \cite{Mies2000FeshbachResWithVaryingBfield, Jaksch2002MolBECFromMott, Busch1998TwoAtomsInTrap}. The free-bound Rabi frequency depends on the Franck-Condon factor as $\Omega_{\mathrm{FB}}\propto \mathrm{FCF_{FB}}\propto \int_{0}^{\infty} \psi_{e}^*(r) \psi_{a}(r) r^2 dr$, where $\psi_{e}$ and $\psi_{a}$ are the radial nuclear wavefunctions of the excited bound state and the initial atomic state, see Fig~\ref{fig:IntroPicture}b. $\psi_{a}$ is determined at short inter-particle distances by the molecular potential, and at long distances by the external harmonic confinement. $\psi_{e}$ is non negligible only at short distances, and determined by the excited molecular potential. Therefore the value of $\mathrm{FCF_{FB}}$ is only determined by $\psi_{e}$ and the short-range part of $\psi_{a}$. We vary $\sqrt{\langle n \rangle}$ by changing the external confinement provided by the lattice, which affects strongly the long-range part of $\psi_{a}$. We now need to determine how this change affects both the shape and the amplitude of the short-range part of $\psi_{a}$.

The short-range length scales $l_{a,e}$ of $\psi_{a,e}$ are set by the location of the wavefunction node with largest internuclear distance and here $l_{a,e} < \unit[10]{nm}$. Since in our case the two atoms described by $\psi_a$ are both in the ground state of a lattice well, the total extent $R$ of $\psi_a$ is on the order of the harmonic oscillator size, which here means $R\gtrsim \unit[100]{nm}$. A modification of the wavefunction at large distance by a change in the lattice confinement can affect the shape of the short-range wavefunction through a phase shift of order $\delta_{\phi}\simeq (l_a - a_s) k$, where $a_s=\unit[6.6]{nm}$ is the s-wave scattering length and $k\simeq 1/R$ is the relative wavevector at long distance. Since $R \gg l_a$ and  $ R \gg a_s$, the phase shift $\delta_{\phi}$ is negligible and does not affect the value of the integral $\mathrm{FCF_{FB}}$. The relative wavefunction of two non-interacting particles in the external potential, $\phi_0(r)$, is a good approximation of $\psi_{a}(r)$ for $r \gg l_a$. Since $l_a\ll R$ we can choose $r$ such that $\psi_{a}(r)\simeq \phi_0(r)\simeq \phi_0(0)$. As the external trap is changed, the amplitude at short distance is then simply scaled by a factor $\phi_0(0)\propto \sqrt{\langle n \rangle}$, thus leading to  $\Omega_{\mathrm{FB}}\propto \mathrm{FCF_{FB}}\propto \sqrt{\langle n \rangle}$.

\subsubsection{Light shifts and loss by $L_{\mathrm{FB,BB}}$ light}
\label{subsubsec: Light shifts and scattering from PA light}

We now characterize the parameters $\delta_{\mathrm{FB}}$ and $\delta_{\mathrm{BB}}$, which are the time- and space-dependent light shifts contributing to the two-color detuning $\delta$ that are induced by the free-bound laser $L_{\mathrm{FB}}$ and the bound-bound laser $L_{\mathrm{BB}}$, respectively. These detunings correspond to binding energy changes $-\hbar \delta_{\mathrm{FB,BB}}$ of molecules in state $\vert m \rangle$. We measure $\delta_{\mathrm{FB}}$ ($\delta_{\mathrm{BB}}$) by two-color dark-state spectroscopy for different intensities of $L_{\mathrm{FB}}$ ($L_{\mathrm{BB}}$) at fixed intensity of $L_{\mathrm{BB}}$ ($L_{\mathrm{FB}}$). We obtain $\delta_{\mathrm{FB}} = 2 \pi \times \unit[18.3(6)]{kHz / (W / cm^{2}) }$ and $\delta_{\mathrm{BB}}= - 2 \pi \times \unit[ 10(10)]{kHz / (W / cm^{2}) }$. These values agree with a simple theoretical estimation, which takes into account only transitions toward the optically excited atomic state $\vert a^* \rangle = {^3P_1} (m_J=0)$. Indeed the light shift induced by $L_{\mathrm{FB,BB}}$ on $\vert m \rangle$ is weak because $L_{\mathrm{FB,BB}}$ do not couple this state to any other bound state out of the $\Lambda$ scheme and because the coupling of $\vert m \rangle$ to $\vert a^* \rangle$ has to be scaled by a free-bound FCF. By contrast the light shift on $\vert a \rangle$ is substantial because of the atomic transition $\vert a \rangle - \vert a^* \rangle$, and therefore dominates the light shift of $\delta$. To calculate the light shifts, we note that for our parameters the Rabi frequencies $\Omega^{\mathrm{FF}}_{\mathrm{FB,BB}}$ induced on the atomic transition by $L_{\mathrm{FB,BB}}$ are much smaller than the detunings of the lasers from the atomic transition. Taking also into account that two atoms contribute, the light shift of the binding energy can be approximated by $\delta_{\mathrm{FB}} \approx \hbar \Omega^{\mathrm{FF}^{2}}_{\mathrm{FB}} /2 E_{e} = 2  \pi \times \unit[20.0]{kHz / (W / cm^{2}) }$ and $\delta_{\mathrm{BB}} \approx \hbar \Omega^{\mathrm{FF}^{2}}_{\mathrm{BB}} /2(E_{e} - E_{m})= - 2  \pi \times \unit[11.0]{kHz / (W / cm^{2}) }$, which is close to the measured values.

The coupling lasers $L_{\mathrm{FB}}$ and $L_{\mathrm{BB}}$ also induce time- and space-dependent light shifts $\Delta_{\mathrm{FB}}$ and $\Delta_{\mathrm{BB}}$ on the free-bound transition. We neglect $\Delta_{\mathrm{BB}}$ because $L_{\mathrm{BB}}$ has low intensity. To determine $\Delta_{\mathrm{FB}}$, we perform one-color spectroscopy at several intensities of $L_{\mathrm{FB}}$. We derive $\Delta_{\mathrm{FB}}= \unit[2\pi \times 21(1)]{kHz / (W / cm^{2}) }$. The same reasoning as above shows that the one-color shift is also dominated by the light shift of $\vert a \rangle$ and that $\Delta_{\mathrm{FB}} \approx\delta_{\mathrm{FB}}$, consistent with the measurement.

Significant atom loss is caused by $L_{\mathrm{FB}}$ through off-resonant scattering of photons on the ${^1S_0} - {^3P_1}$ transition, which is the main contribution to $\gamma_a$. We measure this scattering rate by illuminating an atomic sample in a Mott insulator with $L_{\mathrm{FB}}$ detuned from the free-bound resonance by a few MHz. Since the system is in the Lamb Dicke regime, losses occur only through light-assisted inelastic collisions in sites with at least two atoms. We derive an effective natural linewidth for the free-atom transition of $\Gamma_a= \unit[1.90(8)]{\Gamma_{^3{\rm P}_1}}$, where $\Gamma_{^3{\rm P}_1}$ is the natural linewidth of the intercombination line. This is consistent with superradiant scattering, which is expected since the harmonic oscillator length is much smaller than the wavelength of $L_{\mathrm{FB}}$.

\subsubsection{Light shifts from lattice light}
\label{subsubsec: Light shifts from lattice light}

\begin{figure}[tb]
\includegraphics[width=\columnwidth]{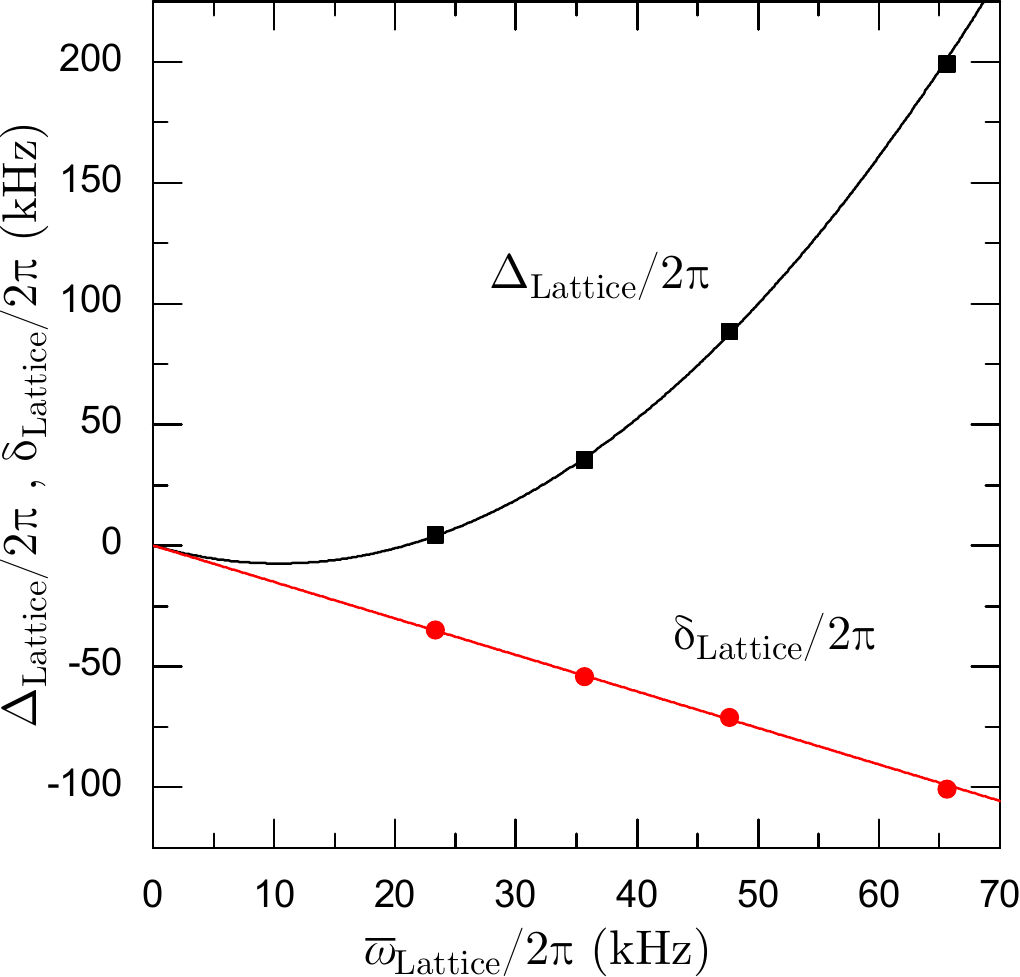}
\caption{\label{fig:BEVSExtConf}(color online) Shift $\delta_{\mathrm{Lattice}}$ of the two-color detuning (red disks) and shift $\Delta_{\mathrm{Lattice}}$ of the free-bound transition (black squares), as a function of the average trap frequency of the central well of the optical lattice. The measurements determine $\delta_{\mathrm{Lattice}}$ ($\Delta_{\mathrm{Lattice}}$) only up to a common offset, which is obtained by fits (curves) described in the text. The error bars of the measurements (i.e. excluding the error in the common offsets) are smaller than the symbol sizes.}
\end{figure}

The lattice light induces static, space-dependent shifts of the two-color detuning ($\delta_{\mathrm{Lattice}}$) and of the free-bound transition ($\Delta_{\mathrm{Lattice}}$), which we now analyze experimentally and theoretically. We measure shifts $\delta_{\mathrm{Lattice}}$ up to a common offset through dark-state spectroscopy for several lattice depths, keeping the lattice well isotropic with $\omega_{x,y,z}$ within $\unit[10]{\%}$ of each other. Similarly we measure shifts $\Delta_{\mathrm{Lattice}}$ through one-color spectroscopy up to a common offset. Both shifts are shown in Fig.~\ref{fig:BEVSExtConf} as a function of the average trap frequency of the central well. As before, the detuning $\delta_{\mathrm{Lattice}}$ corresponds to a change of the molecular binding energy of $ - \hbar \delta_{\mathrm{Lattice}}$.

In the following we identify two independent components of the shifts by separating the two-body problem into its center-of-mass (CM) motion and its relative motion (rel), neglecting mixing terms if present. The first component is induced on the CM by the difference in polarizability of atom pairs and molecules. The second component is induced on the eigenenergies of the relative motion Hamiltonian by the external confinement and enables us to measure the zero-point energy of the lattice wells.

We start by deriving the two components of $\delta_{\mathrm{Lattice}}$. Firstly, the difference in polarizability of state $\vert a \rangle$ and $\vert m \rangle$ leads to a differential shift $\delta^{\mathrm{CM}}_{\mathrm{Lattice}}\propto \left(2\alpha_{ \left(^1S_0\right)} - \alpha_m \right) I_\mathrm{Lattice}$ as the external potentials experienced by the CM differ, where $\alpha_{ \left(^1S_0\right)}$ is the polarizability of a ground-state atom and $\alpha_{m}$ the polarizability of a molecule in state $\vert m \rangle$. Since the polarizability of weakly-bound molecules for far detuned light is close to the sum of the atomic polarizabilities of the constituent atoms we expect this shift to be small \cite{Clark2015QDynamicsOptFeshbachRes, Cetina2015DecoherencePolaron}. Secondly, the external confinement induces a differential shift on the eigenenergies of the relative motion Hamiltonian, leading to $\delta^{\mathrm{rel}}_{\mathrm{Lattice}}$. The relative motion component of state $\vert a \rangle$ occupies the ground state of the lattice well potential with energy $E = 3\, \hbar\overline{\omega}_{\mathrm{Lattice}}/2  \propto I_{\mathrm{Lattice}}^{1/2}$, where $\overline{\omega}_{\mathrm{Lattice}}/2 \pi = (\omega_{x}+\omega_{y}+\omega_{z})/6 \pi$ is the average trap frequency. By contrast, the energy of the relative motion component of state $\vert m \rangle$ is almost insensitive to the external confinement because the corresponding Condon point sets a volume scale which is less than $\unit[0.1]{\%}$ of the ground-state oscillator volume. The light shift $\delta^{\mathrm{rel}}_{\mathrm{Lattice}}$ is therefore dominated by the behavior of $\vert a \rangle$, hence $\delta^{\mathrm{rel}}_{\mathrm{Lattice}}\propto I_{\mathrm{Lattice}}^{1/2}$. In contrast to $\delta^{\mathrm{CM}}_{\mathrm{Lattice}}$ this shift is even present for $\alpha_{m}=2\alpha_{\left(^1S_0\right)}$. We neglect the density dependent interaction shift $\delta^{\mathrm{Coll}}_{\mathrm{Lattice}} \propto n \propto I_\mathrm{Lattice}^{1/4}$, because in our case it is small compared to $\delta^{\mathrm{rel}}_{\mathrm{Lattice}}$. The total light shift is then given by $\delta_{\mathrm{Lattice}}=\delta^{\mathrm{rel}}_{\mathrm{Lattice}} + \delta^{\mathrm{CM}}_{\mathrm{Lattice}} = a_1 \,I_{\mathrm{Lattice}}^{1/2} + b_1 \,I_\mathrm{Lattice}$, with $a_1,b_1 $ being constants. We can thus distinguish the two contributions because of their different scaling with intensity.

The magnitude of the two components of $\delta_{\mathrm{Lattice}}$ can be determined from our experimental data. We fit the data as a function of $\overline{\omega}_{\mathrm{Lattice}} \propto I_{\mathrm{Lattice}}^{1/2}$, i.e. $\delta_{\mathrm{Lattice}} = a_2 \overline{\omega}_{\mathrm{Lattice}} + b_2 \,  {\overline{\omega}^2_{\mathrm{Lattice}}}$, using also a common offset to all $\delta_{\mathrm{Lattice}}$ data points as fit parameter. The fit gives $a_2=1.5(3)$, which is consistent with the expected zero-point energy shift $3 \hbar\,\overline{\omega}_{\mathrm{Lattice}}/2$. The fit result for $b_2$ yields an upper bound for the relative polarizability variation between atom pair and molecule of $\vert (2\alpha_{ \left(^1S_0\right)}-\alpha_{m})/2\alpha_{a}\vert < \unit[1]{\%}$. By assuming $\delta^{\mathrm{rel}}_{\mathrm{Lattice}}$ to dominate the shift and refitting while keeping $b_2=0$, we obtain $a_2 = 1.50(6)$. This measurement directly determines the variation of the harmonic oscillator zero point energy with trap frequency and was possible for two reasons: first, the negligible difference in polarizability between atom pairs and molecules and, second, the high precision achievable with dark-state spectroscopy.

\begin{table}[tb]
\caption{Relevant parameters for molecule production.}
\begin{ruledtabular}
\begin{tabular}{c c c c}
Parameter & Units & Experiment & Theory \\
\hline
\noalign{\smallskip}
$E_m/h$ & $\unit{MHz}$ & $644.7372(2)$ & - \\
$E_e/h$ & $\unit{MHz}$ & $228.38(1)$ & - \\
$\Omega_{\mathrm{FB}}$  & $\unit{ \frac{kHz \, cm^{3/2}}{\sqrt{W/cm^{2}}}}$ & 2$\pi \times 6.0(1) \, 10^{-7}$ & -\\
$\Omega_{\mathrm{BB}}$ & $ \unit{\frac{kHz}{\sqrt{W/cm^{2}}}}$ & $2\pi \times 234(5)$ & -\\
$\Gamma_e$ & $\unit{kHz}$ & $2\pi \times 17.0(1.5)$ & $> 2\pi \times 14.8$\\
$\Gamma_a/ \Gamma_{^3{\rm P}_1}$ & -  & $1.90(8)$ & $2$\\
$\tau_{\mathrm{m}}$ & $\unit{ms}$ & $6(2)\times t_{\mathrm{tunnel}}$ & $\propto t_{\mathrm{tunnel}}$\\
$\gamma_m^{\mathrm{FB}}$ & $\unit{\frac{Hz}{W/cm^{2}}}$ & $2\pi \times 11(1)$ & -\\
$\gamma_m^{\mathrm{COMP}}$ & $\unit{\frac{Hz}{W/cm^{2}}}$ & $2\pi \times 60(6)$ & -\\
$\delta_{\mathrm{FB}}$ &  $\unit{\frac{kHz}{W/cm^{2}}}$ & $+ 2 \pi \times  18.3(6)$ & $+ 2 \pi \times 20.0$\\
$\delta_{\mathrm{BB}}$ &  $\unit{\frac{kHz}{W/cm^{2}}}$ & $- 2 \pi \times 10(10)$ & $- 2 \pi \times 11.0$\\
$\Delta_{\mathrm{FB}}$ &  $\unit{\frac{kHz}{W/cm^{2}}}$ & $+2 \pi \times 21(1)$ & $+2 \pi \times 20.0$\\
$\delta_{\mathrm{Lattice}}$ &  $\overline{\omega}_{\mathrm{Lattice}}$ & $-1.50(6)$ &  $-1.5$\\
$\Delta^{\mathrm{CM}}_{\mathrm{Lattice}}$ & $\unit{\frac{Hz}{W/cm^{2}}}$  & $2 \pi \times 2.13(2)$ &  $2 \pi \times 2.98$\\
$\Delta^{\mathrm{rel}}_{\mathrm{Lattice}}$ & $\overline{\omega}_{\mathrm{Lattice}}$  & $-1.4(5)$ &  $-1.5$\\
$\tau_{\mathrm{DarkState}}$ & $\unit{ms}$ & $2.1(2)$ & -\\
\end{tabular}
\end{ruledtabular}
\label{table:ParametersMI}  
\end{table}

We finally analyze the shift $\Delta_{\mathrm{Lattice}}$ of the free-bound transition. Analogous to $\delta^{\mathrm{CM}}_{\mathrm{Lattice}}$, the centre-of-mass component of the shift is given by $\Delta^{\mathrm{CM}}_{\mathrm{Lattice}} \propto \left( 2 \alpha_{ \left(^1S_0\right)} - \alpha_e \right) I_\mathrm{Lattice}$, where now the molecular state is $\vert e \rangle$ with polarizability $\alpha_e$. Again assuming that the molecular polarizability is the sum of atomic polarizabilities, we have $\Delta^{\mathrm{CM}}_{\mathrm{Lattice}} \propto \left( 2\alpha_{  \left(^1S_0\right)} - \left( \alpha_{ \left(^3P_1\right)} + \alpha_{ \left(^1S_0\right)} \right) \right)  I_\mathrm{Lattice} =\left( \alpha_{ \left(^1S_0\right)} - \alpha_{ \left(^3P_1 \right)}  \right) I_\mathrm{Lattice}$. In contrast to before this shift is not small since the difference in polarizability of the $^1S_0$ and $^3P_1$ ($m_J=0$) states at the lattice wavelength is significant. A smaller shift is induced by changing lattice potentials on the CM zero point energy difference between $\vert a \rangle$ and $\vert e \rangle$, which we neglect. The relative motion component of the shift $\Delta^{\mathrm{rel}}_{\mathrm{Lattice}}$ is given by $\approx - 3 \,\overline{\omega}_{\mathrm{Lattice}}/2$, since also here the relative motion energy of molecular state $\vert e \rangle$ is barely influenced by the external potential. The two components add to the one-color detuning shift $\Delta_{\mathrm{Lattice}}=\Delta^{\mathrm{rel}}_{\mathrm{Lattice}} + \Delta^{\mathrm{CM}}_{\mathrm{Lattice}}= a_3 \,\overline{\omega}_{\mathrm{Lattice}} + K\,\left( \alpha_{ \left(^1S_0\right)} - \alpha_{ \left(^3P_1\right)}  \right) \overline{\omega}_{\mathrm{Lattice}}^2$, where $a_3$ is a free parameter, $K= 3 m \lambda_{\mathrm{Lattice}}^2/(4 \pi h \alpha_{ \left(^1S_0\right)})$, $m$ the mass of $^{84}$Sr, and $\alpha_{ \left(^1S_0\right)}=-\unit[234]{a.u.}$ (atomic units, here $\unit[1]{a.u.}=4\pi \epsilon_0 a_0^3$, where $\epsilon_0$ is the vacuum permittivity and $a_0$ the Bohr radius). Using $a_3$, $\alpha_{ \left(^3P_1\right)}$, and a common offset to all data points as parameters, we fit this function to the data and retrieve $\alpha_{ \left(^3P_1\right)}=-\unit[188(2)]{a.u.}$ and $a_3=-1.4(5)$. Since we experimentally vary the intensity for the three lattice beams together, we expect to obtain the mean value of the polarizability for the state $^3P_1$ ($m_J=0$) calculated for the three different polarizations of the electric field for the three beams, which is $-\unit[170.4]{a.u.}$ \cite{Boyd2007Thesis}, roughly $\unit[10]{\%}$ different from the experimental value. This deviation might be explained by our model neglecting the anisotropy in the trap frequencies.

\subsection{STIRAP}
\label{subsec:MI STIRAP}

We now apply STIRAP to our Mott insulator sample. We prove the association of molecules and measure the molecule association efficiency after optimization of the relevant parameters. We compare this efficiency with our theoretical model and find that dynamic light shifts of the binding energy limit the current scheme. This limit will be overcome in Sec.~\ref{subsec:Light-shift compensation}.

In order to demonstrate the production of molecules by STIRAP and to obtain a quantitative measurement of the single-path STIRAP efficiency, we apply a STIRAP cycle, see Sec.~\ref{sec:Overview}. Between aSTIRAP and dSTIRAP, we selectively remove all remaining atoms with a pulse of light resonant with the atomic ${^1S_0} - {^1P_1}$ transition. Figure \ref{fig:STIRAPCOMBO}c shows the intensity profile used for the STIRAP lasers $L_{\mathrm{FB,BB}}$ and the push pulse. The push pulse doesn't affect molecules because their binding energy is much bigger than the linewidth of the transition $\Gamma_{^1{\rm P}_1}\simeq \unit[2\pi \times 30]{MHz}$. Similarly, since absorption imaging is also performed using this transition, only atoms are imaged. Reappearance of atoms on images taken after the full STIRAP cycle is the experimental signature for the presence of molecules after the aSTIRAP. Assuming an equal efficiency for aSTIRAP and dSTIRAP, the single-path STIRAP efficiency is $\eta= \sqrt{N_{f}/N_{i}}$, where $N_{i}$ and $N_{f}$ are the atom numbers in doubly-occupied sites before and after the STIRAP cycle, respectively.

We optimize the STIRAP sequence by maximizing the number of Sr atoms retrieved after two STIRAP cycles, by varying independently the relevant parameters ($\Omega_{\mathrm{FB}}$, $\Omega_{\mathrm{BB}}$, $T$ and $\langle n \rangle$). We choose to optimize the atom number after two cycles in order to reduce the influence of the inhomogeneous lattice light shift $\delta_{\mathrm{Lattice}}$ on the optimization result. During the first aSTIRAP only atoms on a subset of sites with similar light shift are successfully associated. The push pulse removes all remaining atoms, such that after dSTIRAP we are left with a sample of atom pairs on sites with similar lightshift. This purification of the sample is also evident when comparing the width of one-color PA spectra taken before and after the first STIRAP cycle, see Sec.~\ref{subsubsec:  Inhomogeneous light shifts from lattice light}. The second STIRAP cycle is used to measure the efficiency of STIRAP on this more homogeneous sample. The optimized parameters are reported in Tab.~\ref{table:ParametersSTIRAP} in the column labeled NO-COMP. The best single-path STIRAP efficiency is $\eta_{\mathrm{exp}}= \unit[53.0(3.5)]{\%}$, which represents a considerable improvement compared to our previous work \cite{Stellmer2012Sr2Mol}. This improvement is made possible by a longer molecule lifetime, see Sec.~\ref{subsubsec: Molecule lifetime}.

To compare the performance of the experiment with the theoretical expectation, we model the aSTIRAP using Eq.~(\ref{MI Model}). This model requires two not yet determined parameters, $\gamma_e$ and $\gamma_m$, which can only be characterized employing STIRAP. Their measurement will be discussed in Sec.~\ref{subsubsec: Determination of gammae} and \ref{subsubsec: Molecule lifetime}, respectively. Taking these and all previously determined parameters together with the pulse shape as input for the model, we predict that aSTIRAP has an efficiency of $\eta_{\mathrm{theory}}= \unit[55(5)]{\%}$, which is consistent with our measurements.

\begin{table}[tb]
\caption{Optimized parameters used for STIRAP, without compensated two-color detuning (NO-COMP) or with (COMP).}
\begin{ruledtabular}
\begin{tabular}{c c c c}
Parameter & Units & NO-COMP & COMP\\
\hline
$\Omega_{\mathrm{FB}}$ & $\unit{kHz}$ & $2\pi \times 32(2)$ & $2\pi \times 38(2)$\\
$\Omega_{\mathrm{BB}}$& $\unit{kHz}$ &  $2\pi \times 300(10)$ & $2\pi \times 107(3)$\\[1ex]
 $A= \gamma_{e} \delta_{\mathrm{FB}}/\Omega^{2}_{\mathrm{FB}}  $ & - & $-1.4(3)$ & $0.0(1)$\\[1ex]
$\Delta \omega_{\mathrm{COMP}}$ & $\unit{MHz}$ & - & $2\pi \times 197$\\[1ex]
$T_{\mathrm{pulse}}$ & $\unit{\mu s}$ & $400$ & $400$\\
$n_{\mathrm{peak}}$ & $\unit{cm^{-3}}$ & $4.6 \times 10^{15}$ & $4.6 \times 10^{15}$\\
$\tau_{\mathrm{m}}$ & $\unit{ms}$ & $>10^{5}$ & $>10^{5}$ \\
$\gamma_{e}$ & $\unit{kHz}$ &  $2\pi \times 44(13)$ &  $2\pi \times 44(13)$\\
$\tau_{\mathrm{DarkState}}$ & $\unit{ms}$ & $2.1(2)$ & $2.1(2)$\\
$\eta$ & - & $\unit[53.0(3.5)]{\%}$ & $\unit[81(2)]{\%}$\\
\end{tabular}
\end{ruledtabular}
\label{table:ParametersSTIRAP}
\end{table}

In order to discriminate whether the main limitation to the STIRAP efficiency is the adiabaticity of the sequence or the lifetime of the dark state we examine the criterion $ \vert A \vert \, \alpha = \vert\delta_m\vert T \gg \pi / \sqrt{C} $ derived from Eq.~\ref{STIRAP Eff} for $\delta \approx\delta_{\mathrm{FB}} ( \sin(\theta)^2 - 1/2 )$ and $\Omega_m \approx \Omega_{\mathrm{FB}}\simeq \Omega_{\mathrm{BB}}$. The latter condition is not fulfilled in the experiment. However, the population transfer happens mainly in the time interval during which $\Omega_{\mathrm{FB}}$ and $\Omega_{\mathrm{BB}}$ are of the same order of magnitude, which makes the criterion approximately valid. We obtain $T= \unit[400]{\mu s} \gg \pi / ( \sqrt{C} \delta_{\mathrm{FB}} ) = \unit[88]{\mu s}$, suggesting that we are in the regime where the main loss mechanism is the dissipation from the finite lifetime of the dark state due to $\delta_{\mathrm{FB}} \neq 0$.

\subsection{STIRAP with light-shift compensation}
\label{subsec:Light-shift compensation}

In order to improve the molecule creation efficiency we need to increase the dark-state lifetime. This lifetime is proportional to $A^{-2}$, where $A= \delta_m/ \tilde{\gamma} \approx \gamma_e \delta_{\mathrm{FB}}/ \Omega_{\mathrm{FB}}^2$, and efficient operation is ensured in the adiabatic regime only if $ \vert A \vert\ll 1$. We cannot decrease $\vert A \vert$ by increasing the intensity of $L_\mathrm{FB}$, because $\delta_\mathrm{FB} \propto \Omega_{\mathrm{FB}}^2$. In the following, we show how to compensate the shift $\delta_{\mathrm{FB}}$ both in time and space by using an additional ``compensation" laser beam. After optimization, the compensation scheme allows us to reach an efficiency of $\eta_{\mathrm{exp}}= \unit[81(2)]{\%}$.

\begin{figure}[tb]
\includegraphics[width=\columnwidth]{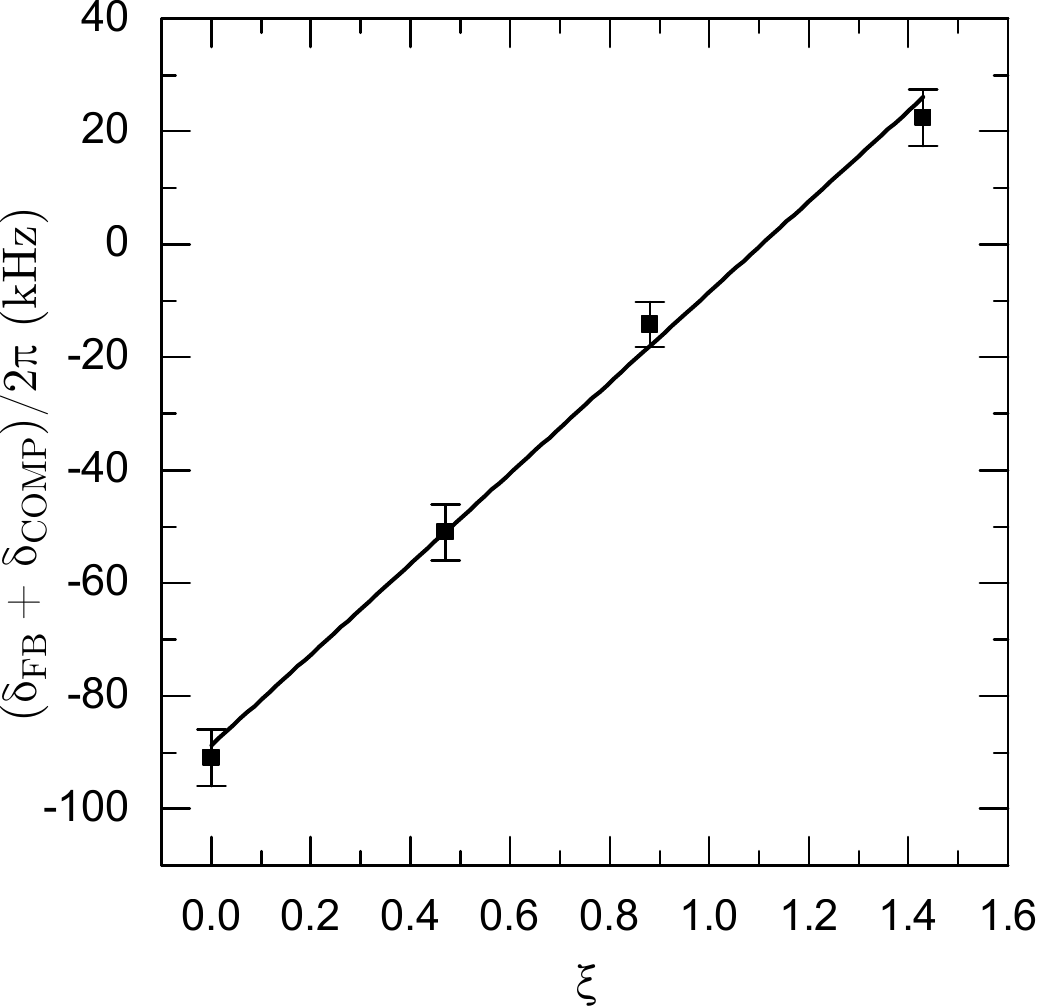}
\caption{\label{fig:LightShiftCompensation} Light shift $\delta_{\mathrm{FB}}+\delta_{\mathrm{COMP}}$ induced by the free-bound laser $L_{\mathrm{FB}}$ and the compensation beam $L_{\mathrm{COMP}}$ as a function of the compensation level $\xi$. The solid line is a linear fit to the data.}
\end{figure}

\subsubsection{Compensation beam}
\label{subsubsec:Compensation beam}

As explained in Sec.~\ref{subsubsec: Light shifts and scattering from PA light}, the two-color detuning shift $\delta_{\mathrm{FB}}$ is dominated by the light shift of $\vert a \rangle$ in presence of L$_{\mathrm{FB}}$ because of the atomic ${^1S_0} - {^3P_1}$ transition, from which L$_{\mathrm{FB}}$ is red detuned by only $\vert \Delta \omega_e \vert= (E_e + \Delta)/ \hbar \approx E_e/ \hbar = \unit[2\pi \times 228]{MHz}$.  The resulting shift $\delta_{\mathrm{FB}} \approx \hbar \Omega^{\mathrm{FF}^{2}}_{\mathrm{FB}} /2 E_e \propto I_{\mathrm{FB}}/\Delta \omega_e$ can be exactly cancelled by superimposing an additional compensation laser field $L_{\mathrm{COMP}}$ with L$_{\mathrm{FB}}$, creating the shift $\delta_{\mathrm{COMP}}=-\delta_{\mathrm{FB}}$ \cite{Yatsenko1997STIRAPwithStarkShiftComp}. To cancel the shift, $L_{\mathrm{COMP}}$ has to be detuned by $\Delta \omega_{\mathrm{COMP}}$ to the blue of the atomic transition (see Fig.~\ref{fig:IntroPicture}a) and must have an intensity $\frac{I_{\mathrm{COMP}}}{I_{\mathrm{FB}}} = \vert \frac{\Delta \omega_{\mathrm{COMP}}}{\Delta \omega_e}\vert$.  This light-shift cancellation technique is similar to the SCRAP method \cite{Yatsenko1999FirstSCRAP}, as it relies on tailoring the light shifts with an off-resonant beam.

\begin{figure}[tb]
\includegraphics[width=\columnwidth]{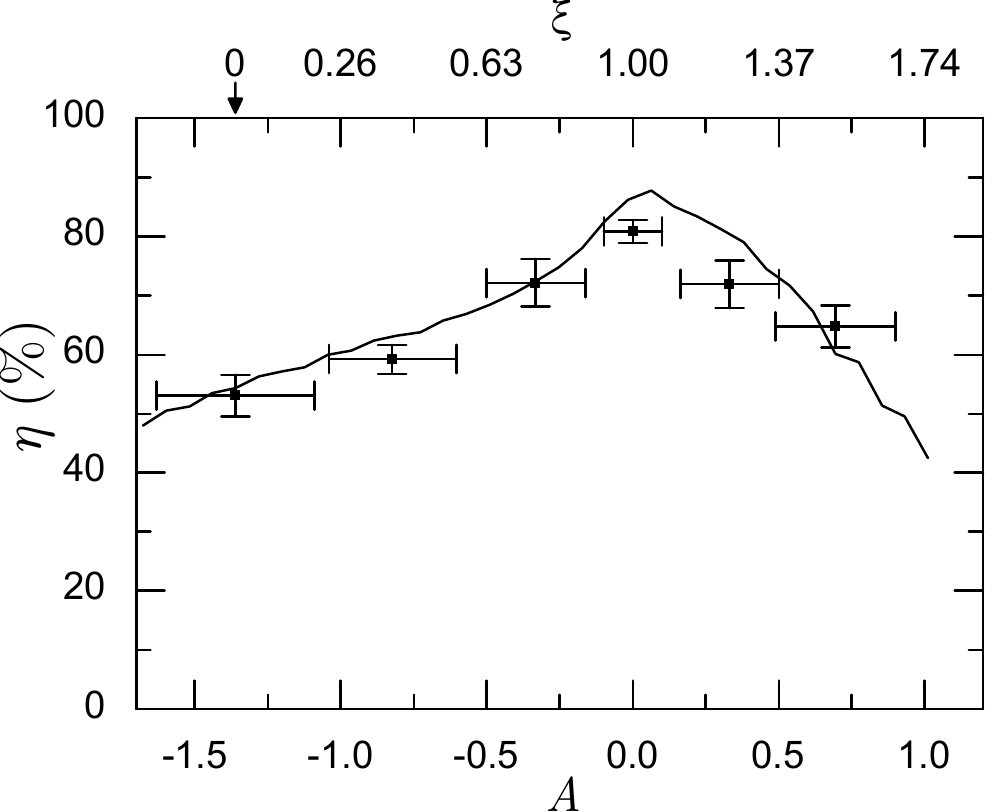}
\caption{\label{fig:EfficiencyVSCompensation} Maximum single-path STIRAP efficiency as a function of the parameter $A= \delta_m/ \tilde{\gamma}$, varied by changing the intensity $I_{\mathrm{COMP}}$ of the compensation beam. Values of the compensation level $\xi$ are given on the top axis for reference. Theoretical values (line) are obtained by simulating the STIRAP process with Eq.~(\ref{MI Model}), using independently determined parameters and not performing any fit. The error bars represent one standard deviation.}
\end{figure}

Since $I_{\mathrm{FB}}$ varies in time during STIRAP the light-shift compensation needs to be dynamic as well \cite{Guerin1998HyperSTIRAPStarkComp}. Since we want to keep $\Delta \omega_{\mathrm{COMP}}$ and $\Delta \omega_e$ constant for convenience, $I_{\mathrm{COMP}}$ has to be varied proportionally to $I_{\mathrm{FB}}$ to always keep the light shift canceled. A simple technical solution to obtain such a coordinated change in intensity is to split the free-bound laser beam into two beams, impose on one of the beams a frequency offset $\vert \Delta \omega_e \vert + \vert \Delta \omega_{\mathrm{COMP}} \vert$ and recombine the beams with exactly the same polarization and spatial mode, before passing through an AOM to control the intensity of both frequency components in common. This composite beam is finally recombined with the bound-bound beam into one single mode fiber, from which PA light is shone on the atoms (see Fig.~\ref{fig:LevelSchemeCompensation}). This setup allows us to compensate the light shift $\delta_{\mathrm{FB}}$ both in time and space.

To validate our technique we measure the light shift $\delta_{\mathrm{FB}}$ induced by $L_{\mathrm{FB}}$ (referenced to the non-shifted extrapolated value for $I_{\mathrm{FB}}= \unit[0]{W/cm^2}$) as a function of the compensation level defined as $\xi= \frac{I_{\mathrm{COMP}}}{I_{\mathrm{FB}}} \vert \frac{\Delta \omega_e}{\Delta \omega_{\mathrm{COMP}}}\vert$, where we expect perfect compensation for $\xi=1$. The compensation level is varied by changing the intensity of the compensation beam before AOM 1 while keeping all other parameters fixed. For this measurement we use $\Delta \omega_{\mathrm{COMP}}=2\pi \times \unit[66]{MHz}$ and an intensity of $I_{\mathrm{FB}}= \unit[5]{W/cm^2}$, which is of the order of the optimum intensity of $L_{\mathrm{FB}}$ for STIRAP. The measured $\delta_{\mathrm{FB}}$ is plotted as a function of $\xi$ in Fig.~\ref{fig:LightShiftCompensation} together with a linear fit from which we obtain $\xi(\delta_{FB}=0)=1.1(1)$, which is consistent with our assumption of the light shift being dominated by the atomic ${^1S_0} - {^3P_1}$ transition. To reduce the off-resonant scattering rate of photons on the atomic transition, we increase $\Delta \omega_{\mathrm{COMP}}$ to $2\pi \times \unit[197]{MHz}$ in all further usages of the compensation beam.

\begin{figure}[tb]
\includegraphics[width=\columnwidth]{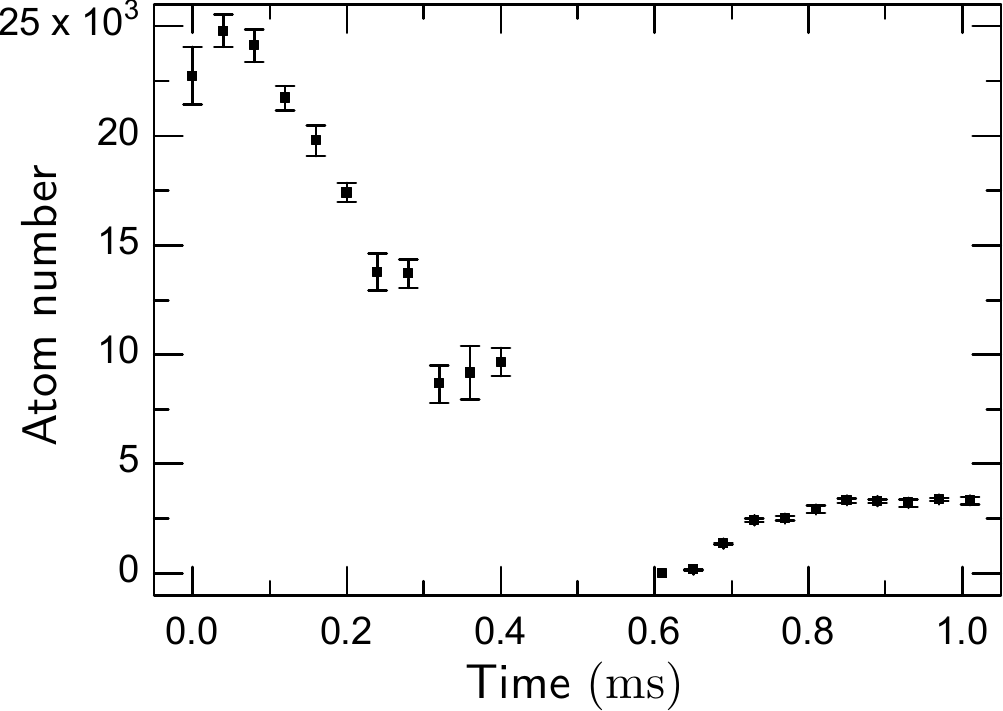}
\caption{\label{fig:FirstSTIRAP} Time evolution of Sr atom number during the first STIRAP cycle, using light-shift compensation. }
\end{figure}

\begin{figure}[tb]
\includegraphics[width=\columnwidth]{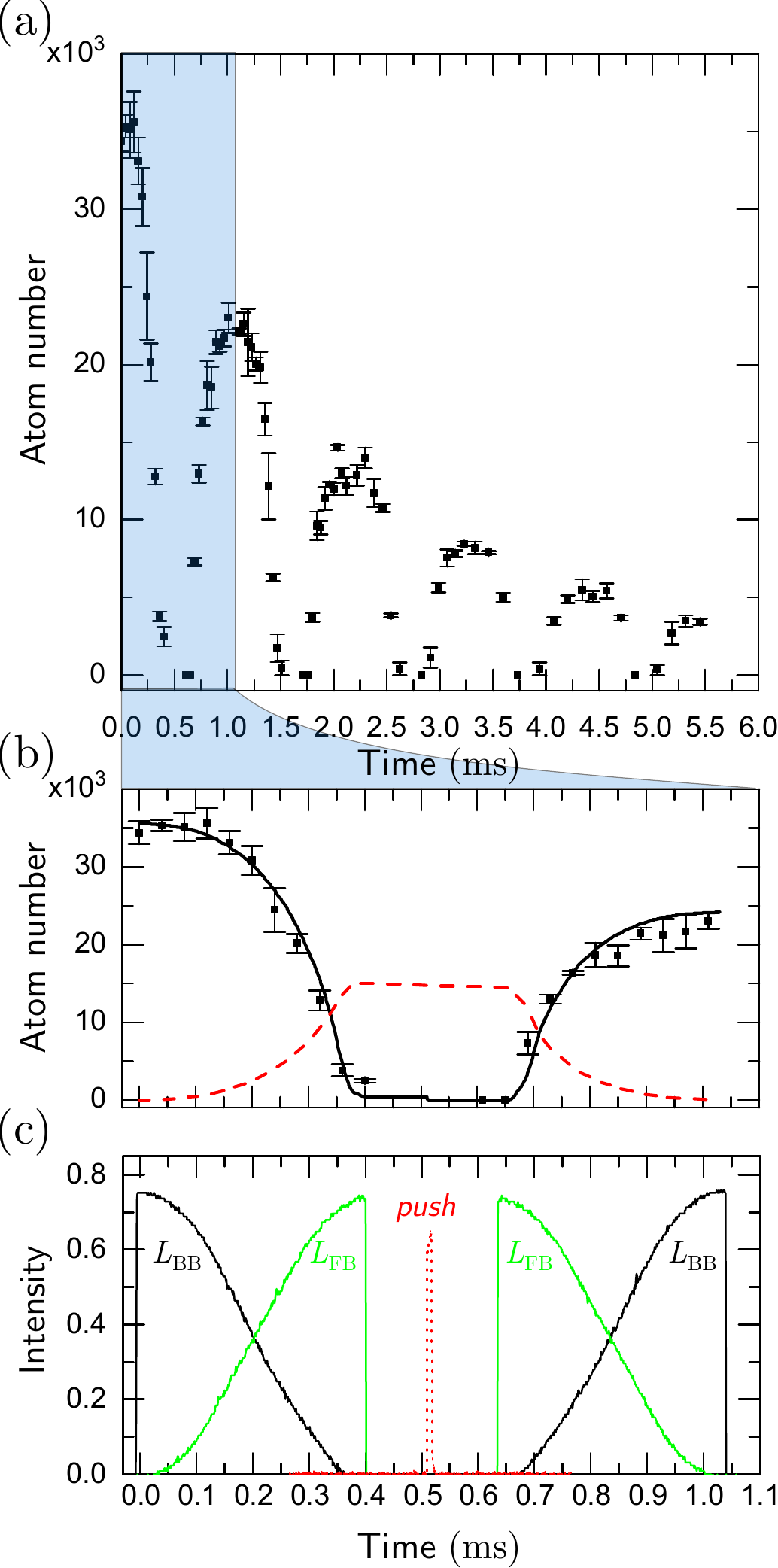}
\caption{\label{fig:STIRAPCOMBO}(color online) (a) Time evolution of the atom number during STIRAP cycles 2 to 5, using light-shift compensation. (b) Number of Sr atoms during the second STIRAP cycle as function of time. The lines are the theory curves based on Eq.~(\ref{MI Model}) with no fitting parameter, representing the atom (black solid line) and molecule (red dashed line) number for the parameters given in Tab.~\ref{table:ParametersSTIRAP}.(c) Intensities of $L_{\mathrm{FB}}$,  $L_{\mathrm{BB}}$ and push pulse beam in arbitrary units. The intensity of $L_{\mathrm{COMP}}$ is not shown.}
\end{figure}

\subsubsection{STIRAP optimization and characterization}
\label{subsubsec: STIRAP optimization}

We characterize the effect of our dynamic optical compensation scheme on the STIRAP efficiency. For several compensation levels $\xi$ we optimize STIRAP and show the efficiency achieved during the second STIRAP cycle in Fig.~\ref{fig:EfficiencyVSCompensation}. The maximum efficiency rises from $\unit[53.0(3.5)]{\%}$ without compensation to $\unit[81(2)]{\%}$ and is obtained for $\xi=1$, corresponding to $A=0.0(1)$. This proves that a strong limitation to the STIRAP efficiency without compensation indeed originates from the space- and time-dependent light shift imposed on the binding energy by $L_{\mathrm{FB}}$ and that our scheme is able to compensate this undesired shift. The optimized parameters, shown in column ``COMP" of Tab.~\ref{table:ParametersSTIRAP}, lead our theoretical model to describe the experimental data well. The model indicates that now, for $\vert A \vert\ll 1$, the limitations to the transfer efficiency are to similar amounts off-resonant scattering of photons from $\vert a \rangle$ and $\vert m \rangle$, and the finite dark-state lifetime resulting, e.g., from residual $\delta_{\mathrm{FB}}$ and $\delta_{\mathrm{BB}}$. All STIRAPs discussed in the remainder of this article use the compensation beam with $\xi= 1$.

We next characterize the evolution of the atom number over several STIRAP cycles, showing the first cycle in Fig.~\ref{fig:FirstSTIRAP} and subsequent cycles in Fig.~\ref{fig:STIRAPCOMBO}a. We observe that the fraction of atoms reappearing after a STIRAP cycle compared to the atom number at the beginning of that cycle is $\sim \unit[65]{\%}$ for all but the first cycle, for which it is only $\unit[14]{\%}$. Ignoring the first STIRAP cycle and assuming equal efficiencies of aSTIRAP and dSTIRAP within each cycle, we find that the single-path STIRAP efficiency is roughly $\unit[80]{\%}$ and constant from the second cycle onwards.

\begin{figure}[tb]
\includegraphics[width=\columnwidth]{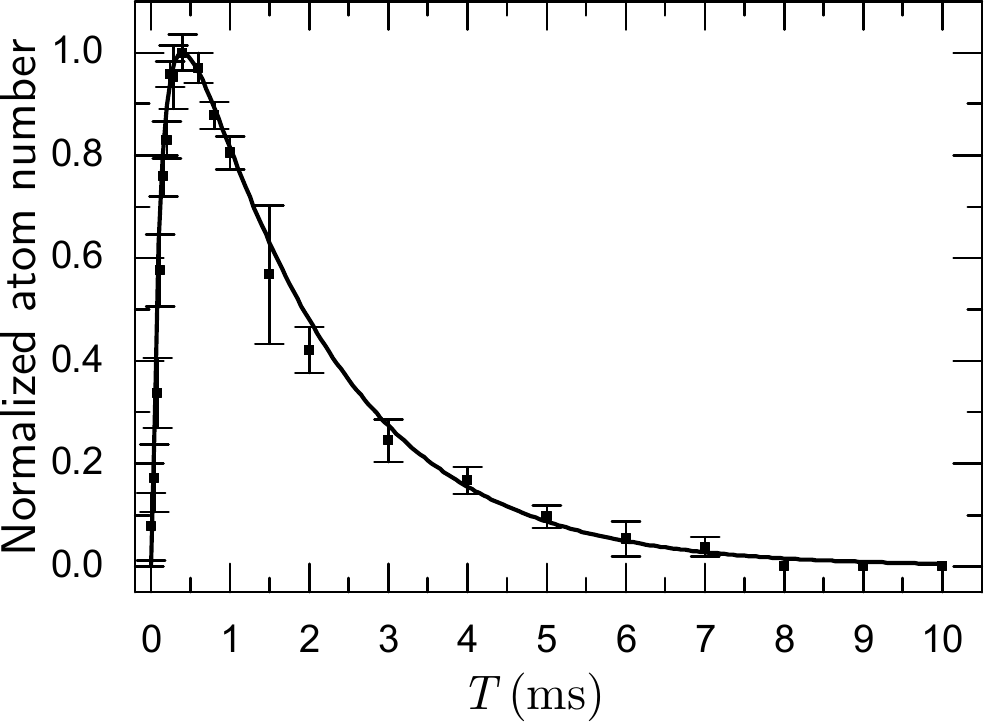}
\caption{\label{fig:RampTime} Number of atoms retrieved after two STIRAP cycles as a function of the pulse time $T$ used for the second STIRAP. The line is a fit using our model, from which we obtain the parameter $\langle \delta \rangle$.}
\end{figure}

The lower fraction of returning atoms of the first cycle can be explained by two effects, the existence of singly-occupied lattice sites and the lattice induced inhomogeneity of the molecular binding energy, which leads to a finite dark-state lifetime and therefore loss. An independent one-color spectroscopy measurement shows that our lattice contains $1.1 \times 10^{5}$ atoms on singly occupied sites, which matches the number of atoms remaining after the first aSTIRAP, letting us conclude that these atoms simply did not have partners to form molecules with. Following the first aSTIRAP these atoms are removed by the push beam pulse. The atom number reduction during the first aSTIRAP of $1.4 \times 10^5$ matches the number of atoms on doubly occupied sites. Assuming an $\unit[80]{\%}$ single-path dSTIRAP efficiency as observed in STIRAP cycles beyond the first, the $3.5 \times 10^{4}$ atoms reappearing after dSTIRAP correspond to $4.4 \times 10^4$ atoms associated into molecules, which is only $\unit[30]{\%}$ of the initial atom number on doubly occupied sites, not $\unit[80]{\%}$ as during later STIRAP cycles. This difference can be explained by the decrease of STIRAP efficiency $\eta$ with increasing two-color detuning $\delta$, characterized in Sec.~\ref{subsubsec: Determination of gammae}, in combination with the inhomogeneous spread of $\delta$ across the sample originating from the lattice light shift, estimated in Sec.~\ref{subsubsec: Inhomogeneous light shifts from lattice light}. Both, $\eta(\delta)$ and the fraction of population on sites with detuning $\delta$, $p(\delta)$, show peaks of similar width. The average STIRAP efficiency can be estimated by averaging $\eta(\delta)$ with weight $p(\delta)$ and is about half of the maximum efficiency, consistent with our observation. This indicates that the inhomogeneous shift of the binding energy by the lattice light shift is the main limitation to the fraction of lattice sites usable for molecule association. In future work this limit could be overcome by increasing the width of the lattice beams, while keeping the lattice depth constant. Here we simply use the first STIRAP cycle to remove atoms on sites with large two-color detuning $\delta$ and atoms on singly-occupied sites, providing us with an ideal sample to study STIRAP.

As first such study we trace the atom number during five consecutive STIRAP cycles in Fig.~\ref{fig:STIRAPCOMBO}a, each identical to the first. The single-path STIRAP efficiency of the second cycle is $\unit[81(2)]{\%}$. All cycles have efficiencies around $\unit[80]{\%}$, from which we can take two conclusions. First, the cleaning of the sample realized by the first STIRAP cycle is enough to decrease the binding energy inhomogeneity to a level that is negligible for the STIRAP efficiency. Second, it proves that our MI sample is not heated significantly, which would have reduced the STIRAP efficiency with each cycle.

Next we study the dependence of STIRAP efficiency on pulse time, by recording the atom number after the second STIRAP cycle, see Fig.~\ref{fig:RampTime}. Thanks to the light-shift compensation scheme, the pulse time for the STIRAP can be as high as a few ms while still resulting in substantial molecule production. For pulse times longer than $\unit[400]{\mu s}$, we observe a decrease of the retrieved atom number on a $1/e$ time of $\sim \unit[2]{ms}$. This decrease can neither be explained by the residual $\delta_{\mathrm{FB}}$ nor by $\delta_{\mathrm{BB}}$, but it can be explained by the finite dark state lifetime of $\tau_{\mathrm{DarkState}} = \unit[2.1(2)]{ms}$ measured in Sec.~\ref{subsubsec:  Inhomogeneous light shifts from lattice light}. This finite lifetime could originate in laser noise or in a small but non-zero static Raman detuning $\langle \delta \rangle$ present during the STIRAP cycle. Fitting our data with the model Eq.~(\ref{MI Model}) using a static Raman detuning as only free parameter we obtain $\langle \delta \rangle = \unit[2\pi \times 2.4(5)]{kHz}$. A similar detuning would explain the observed dark-state lifetime.

\begin{figure}[tb]
\includegraphics[width=\columnwidth]{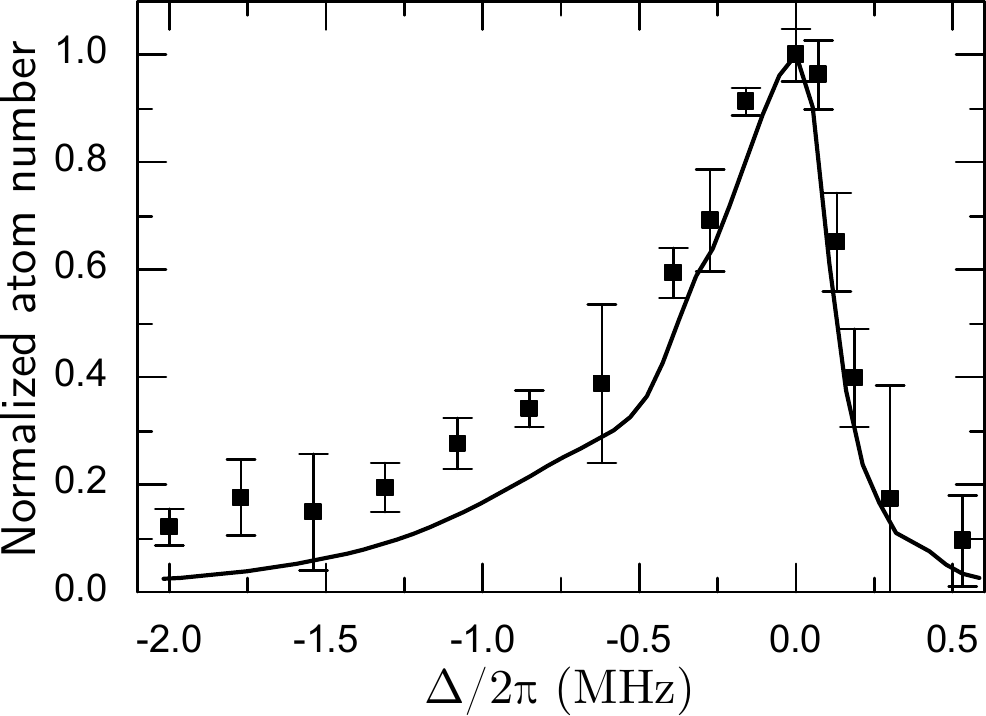}
\caption{\label{fig:UpperStateDetuning} STIRAP efficiency in dependence of the one-color detuning $\Delta$. The data points show the normalized number of atoms retrieved after two STIRAP cycles, depending on the value $\Delta$ used for the second STIRAP. The solid line is derived from Eq.~(\ref{MI Model}) using no fit parameters.}
\end{figure}

\subsubsection{Effect of $\Delta$, $\delta$ on efficiency and determination of $\gamma_e$}
\label{subsubsec: Determination of gammae}

\begin{figure}[tb]
\includegraphics[width=\columnwidth]{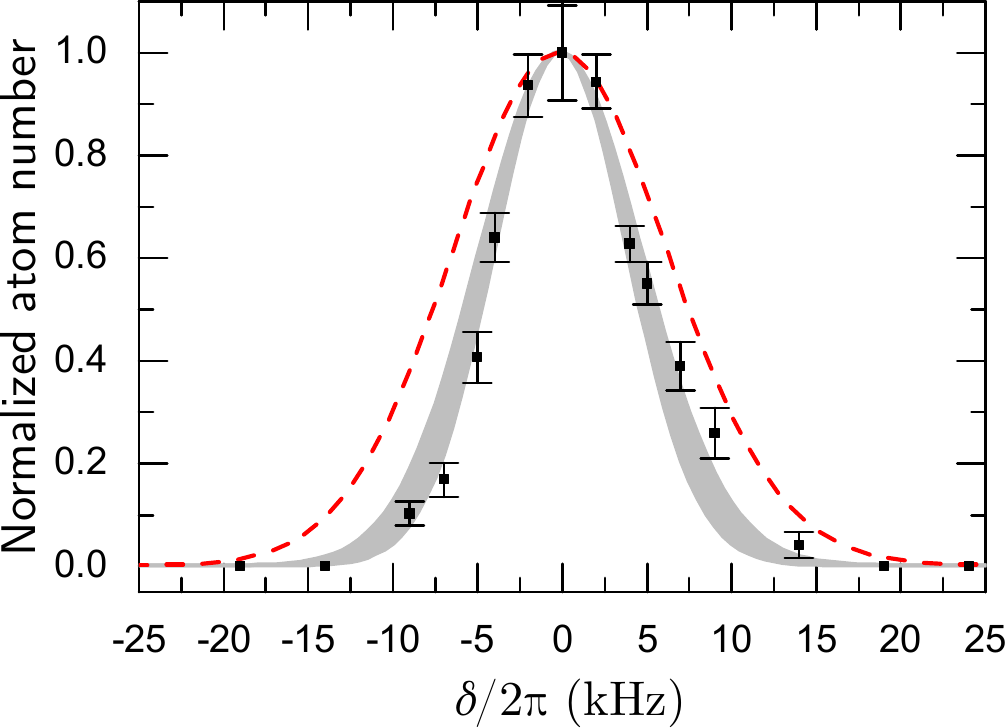}
\caption{\label{fig:RamanLine}(color online) Dependence of STIRAP efficiency on the two-color detuning $\delta$. The data points show the normalized number of atoms retrieved after two STIRAP cycles. The red dashed line is the shape determined by our model for $\gamma_{e}=\Gamma_e= \unit[2\pi \times 17.0(1.5)]{kHz}$, while the grey area spans the widths between $\unit[2\pi \times 30]{kHz}$ and $\unit[2\pi \times 57]{kHz}$, which are the fitted values for the positive-detuning and the negative-detuning side, respectively.}
\end{figure}

We now analyze the efficiency of STIRAP in dependence of the one-color detuning $\Delta$ and the two-color detuning $\delta$, see Fig.~\ref{fig:UpperStateDetuning} and \ref{fig:RamanLine} respectively. To this end, we keep the parameters of the first, purification STIRAP constant, and record the atom number $N_2$ after the second STIRAP for varying $\delta$ while keeping $\Delta=0$, or \textit{vice versa}. The STIRAP efficiency $\eta \propto \sqrt{N_2}$ exhibits peaks around zero detuning, which have a FWHM of $\unit[2\pi \times 0.5]{MHz}$ for $\eta(\Delta)$ and $\unit[2\pi \times 20]{kHz}$ for $\eta(\delta)$. The width of $\eta(\Delta)$ is much bigger than the broadened free-bound linewidth, see Sec.~\ref{subsubsec: Inhomogeneous light shifts from lattice light}, so that the shift due to the inhomogeneous broadening induced by the lattice light on the free-bound transition can be neglected. However, as explained in Sec.~\ref{subsubsec: STIRAP optimization}, the width of $\eta(\delta)$ is comparable to the spread in $\delta$, leading to a lower fraction of atoms in doubly occupied sites being associated during the first aSTIRAP compared to consecutive aSTIRAPs.

From our measurement of $N_2(\delta)$ in Fig.~\ref{fig:RamanLine}, we are able to derive the loss term $\gamma_{e}$ of Eq.~(\ref{MI Model}), which describes the rate at which population is lost from state $\vert e \rangle$ because of coupling to states outside the subspace corresponding to our $\Lambda$ scheme. It can be viewed as an indicator of how open our quantum system is. The value of $\gamma_{e}$ relevant for STIRAP can be higher than the natural linewidth of the free-bound transition, i.e. $\gamma_{e}\geq \Gamma_e$, because of one- or two-photon processes induced by $L_{\mathrm{FB}}$, $L_{\mathrm{BB}}$, and $L_{\mathrm{COMP}}$ that introduce dissipation. To measure $\gamma_e$, we make use of the fact that the STIRAP efficiency is only high for small $A \approx \delta \gamma_e / \Omega_{\mathrm{FB}}^2$, see Sec.~\ref{subsec:Parameter constraints}, which leads to a $\gamma_e$ dependence of the $N_2(\delta)$ peak width. If we assume $\gamma_e=\Gamma_e$, where $\Gamma_e$ is determined in Sec.~\ref{subsubsec: Inhomogeneous light shifts from lattice light}, our model gives the dashed line shown in the figure, which is broader than the observed peak. Moreover, we observe a slight asymmetry in the data that is not reproduced by the model. We therefore fit the model independently to the data corresponding to positive and negative detunings using $\gamma_e$ as the only fit parameter and derive $\gamma_e^+= \unit[2\pi \times 30(5)]{kHz}$ and $\gamma_e^-= \unit[2\pi \times 57(8)]{kHz}$, respectively. In further uses of our model we fix $\gamma_e$ to the average value $\unit[2\pi \times 44(13)]{kHz}$.

\begin{figure}[tb]
\includegraphics[width=\columnwidth]{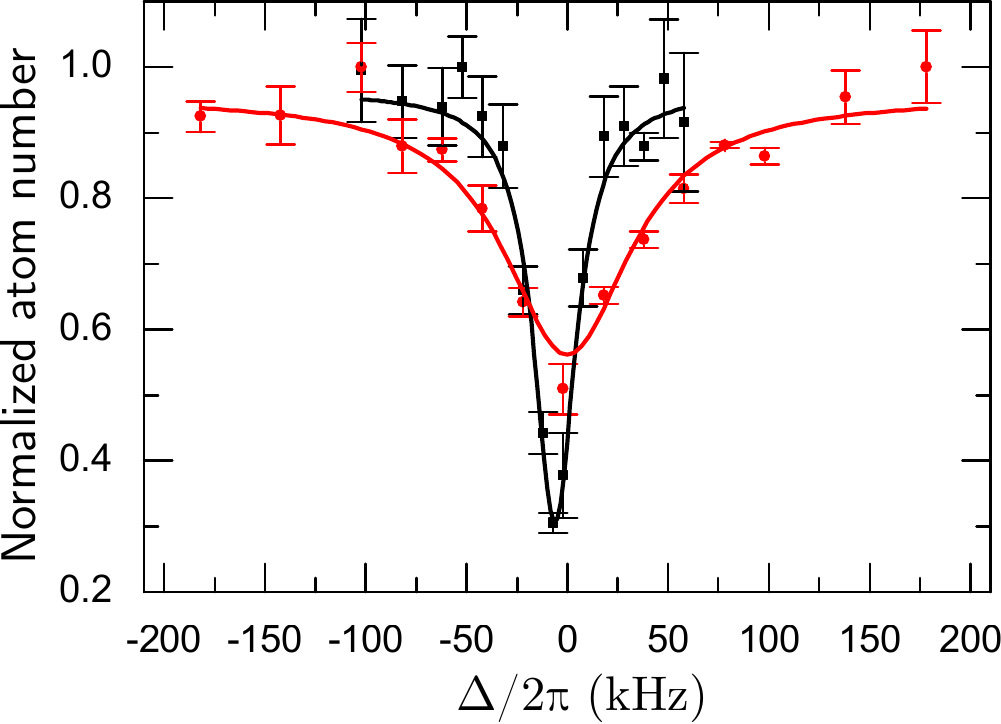}
\caption{\label{fig:1 color Natural linewidth} (color online) One-color molecular line and corresponding Lorentzian fits measured on an atomic sample in the lattice before (red disks) and after (black squares) the first STIRAP cycle. The narrower width of the latter signal is the result of the spectral selection imposed by the STIRAP pulse.}
\end{figure}

\subsection{Sample characterization}
\label{subsec:Sample characterization}

\subsubsection{Inhomogeneous light shifts by lattice light\\ and dark state lifetime}
\label{subsubsec: Inhomogeneous light shifts from lattice light}

In order to help understand the origin of the difference in STIRAP efficiencies between the first STIRAP cycle and the following cycles, we measure the lattice light shifts on the free-bound transition and on the binding energy of $\vert m \rangle$ for both initial and purified samples. We use PA spectroscopy for all measurements, but for the determination of the binding energy spread of the purified sample. Since this spread is below the resolution of our two-color PA spectroscopy, we in this case measure the dark state lifetime and extract an upper bound of the binding energy spread from that.

In order to measure the inhomogeneous broadening induced by the lattice on the free-bound transition, we take one-color PA spectra using a MI sample, either directly after lattice loading or after one STIRAP cycle, see Fig.~\ref{fig:1 color Natural linewidth}. These data sets are modelled by Eqs. (1), where $\gamma_e$ is the fit parameter and all other quantities in the Hamiltonian, except for $\Omega_\mathrm{FB}$, are set to zero. We retrieve $\gamma_e=63(8)\,$kHz and 17.0(1.5)\,kHz for the free-bound transition linewidth before and after STIRAP, respectively. The latter value is consistent with our measurement of the natural linewidth of the molecular transition in a BEC \cite{ToBePublished}, $\Gamma_e = \unit[2\pi \times 19.2(2.4)]{kHz}$, while the former is roughly a factor of 4 wider. The reduction of the apparent linewidth after one STIRAP cycle is a result of the sample purification by the STIRAP cycle. Since both the free-bound transition shift and the binding energy shift are strongly dependent on the shift of state $\vert a \rangle$ by the lattice light, the STIRAP cycle not only reduces the spread in $\delta$, but also the inhomogeneous broadening of the one-color PA line. Deconvolving the measured linewidths with twice the atomic ${^1S_0} - {^3P_1}$ transition $2 \times \Gamma_{^3\mathrm{P}_1} = \unit[2\pi \times 14.8]{kHz}$ \cite{Jones2006RevPA, Reinaudi2012Sr2Mol}, we derive that the inhomogeneous differential light shifts on the free-bound transition of the initial and the final sample are $\unit[2\pi \times 61(10)]{kHz}$ and $\unit[2\pi \times 8(3)]{kHz}$, respectively.

In order to measure the inhomogeneous broadening induced by the lattice on the binding energy of $\vert m \rangle$, we perform two distinct measurements depending on the sample. We perform two-color spectroscopy on the initial sample using low intensity of $L_{\mathrm{FB,BB}}$, which ensures negligible light shifts from the PA light. The width of the dark resonance is then $\Delta \omega_{\mathrm{DarkState}} = \unit[2\pi \times 18(4)]{kHz}$. To deduce an upper bound on the binding energy spread of the purified sample, we measure the dark-state lifetime. We prepare a roughly equal superposition of atomic and molecular state by applying only half of the second aSTIRAP pulse. We wait for a variable time while holding $L_{\mathrm{FB,BB}}$ at constant intensity, and finally apply the second half of the dSTIRAP. The measured lifetime is $\tau_{\mathrm{DarkState}} = \unit[2.1(2)]{ms}$, and the experimental data agree with our model when adding a static two-color detuning of $\langle \delta \rangle = \unit[2\pi \times 2.2(2)]{kHz}$. This value is an upper bound of the binding energy spread induced by the lattice on the purified sample.

The ratio of the free-bound transition linewidth and the two-color detuning spread is consistent with the light-shift calibration, both before and after STIRAP (see Tab.~\ref{table:ParametersMI}). Using this calibration, we conclude that the initial sample was populating lattice sites located off axis by up to $\simeq \unit[30]{\%}$ of the lattice beam waist.

\begin{figure}[tb]
\includegraphics[width=\columnwidth]{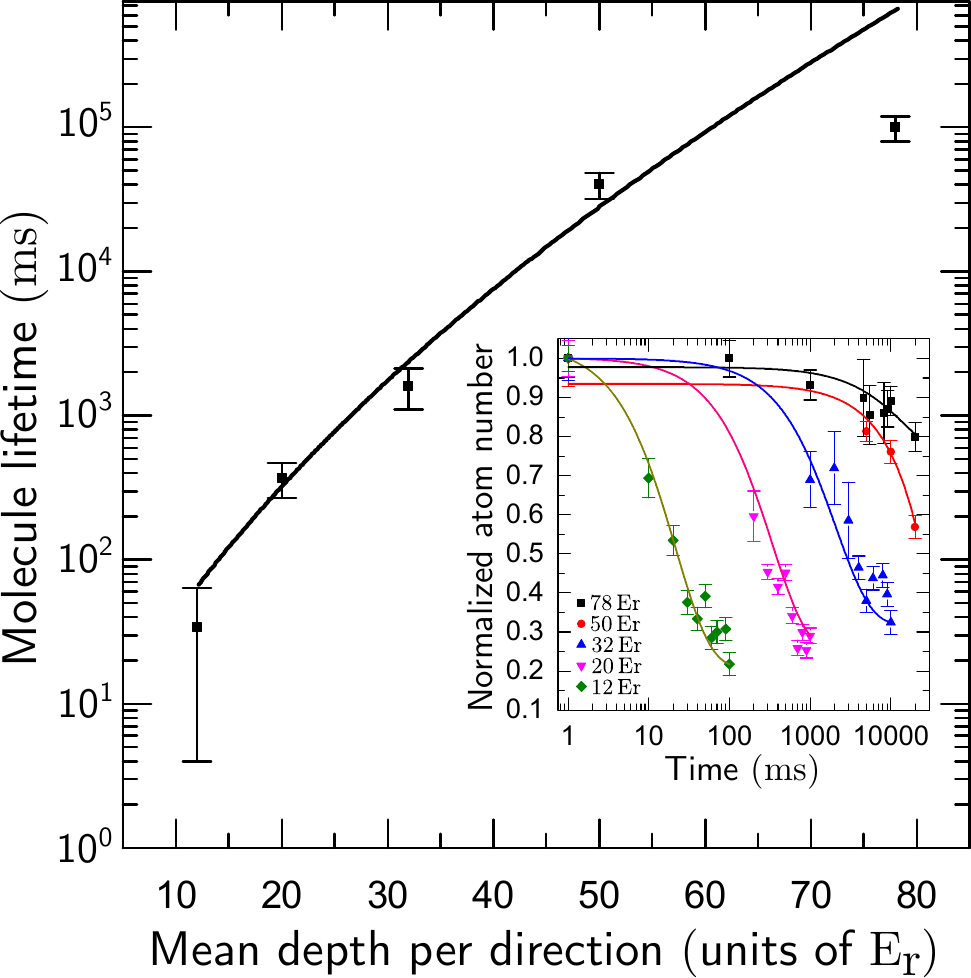}
\caption{\label{fig:MoleculeLifetime} (color online) Molecule lifetime as function of the lattice depth averaged over the three directions in the central region of the lattice. The depth is that of the potential experienced by the molecules, expressed in units of recoil energies of the molecule $\mathrm{E_r} /h  = \unit[8.4]{kHz}$. Inset: atom number decay as a function of hold time between association and dissociation STIRAP pulses for different lattice depths.}
\end{figure}

\subsubsection{Molecule lifetime}
\label{subsubsec: Molecule lifetime}

The molecules produced in our experiment have a finite lifetime as a result of two loss mechanisms, inelastic collisions with other molecules in the lattice and dissipation caused by optical coupling of state $\vert m \rangle$ to states lying outside our $\Lambda$ scheme. Thus, the lifetime is given by $\tau_m=1/ \gamma_m= 1/( \gamma_m^{\mathrm{coll}}+ \gamma_m^{\mathrm{opt}})$, where $\gamma_m^{\mathrm{coll}}$ is the decay rate for collisional losses and $\gamma_m^{\mathrm{opt}}$ is the scattering rate for optically induced losses.

We measure the molecule lifetime $\tau_m^{\mathrm{coll}}=1/ \gamma_m^{\mathrm{coll}}$ that results from inelastic collisions by varying the hold time between the push pulse and the dSTIRAP at several lattice depths. During these measurements the lattice well potential is kept isotropic with $\omega_{x,y,z}$ within $\unit[10]{\%}$ from each other. The molecule lifetime depends strongly on the lattice depth, see Fig.~\ref{fig:MoleculeLifetime}. The observed change of four orders of magnitude over the lattice depth is consistent with loss of molecules by tunneling to neighboring sites followed by effectively instantaneous chemical reaction with the molecule already present on that site. Collisions between molecules and atoms do not play a role here since all atoms have been removed by the push pulse prior to lowering the lattice depth. Under these assumptions the lifetime is given by $\tau_m^{\mathrm{coll}} = D \times t_{\mathrm{tunnel}}$, where $D$ is a constant and $t_{\mathrm{tunnel}}=h / \sum_{i}2J_{i}$ is the total tunneling time, which depends on the tunneling rates along the 3 lattice directions $J_{1,2,3}$ \cite{Jaksch1998AtomsInLattices, Zwerger2003MottHubbard}. We fit the experimental data with this single-parameter model and retrieve $D=6(2)$. We observe that the lifetime measured for our deepest lattice are lower than predicted by our model. This can be explained by loss through scattering of lattice photons becoming dominant over the low inelastic collision loss, since here the lattice light intensity is high and the tunnel time scale long.

For the deepest lattice the molecule lifetime is $\unit[100(20)]{s}$, representing a major improvement compared to our previous paper, where the lifetime was about $\unit[60]{\mu s}$ \cite{Stellmer2012Sr2Mol}. The main difference between this work and our previous work is the wavelength of the light used for the realization of the optical lattice. In our previous work, the lattice laser source was a Coherent Verdi at 532\,nm wavelength, whereas here we use an Innolight Mephisto MOPA at 1064\,nm. By shining 532-nm light derived from another Verdi laser on our sample, we confirm a strong reduction of the molecule lifetime within a factor of 10 of what was measured before. An explanation for the reduced lifetime is that the 532-nm light is close to molecular lines, thus inducing inelastic processes.

The second loss mechanism arises from optical couplings of molecules toward states outside our $\Lambda$ scheme. One contribution is the scattering of lattice photons, as mentioned above. Other significant contributions arise from $L_{\mathrm{FB,COMP}}$, whereas the low intensity $L_{\mathrm{BB}}$ plays a minor role. We measure the scattering time $\tau_m^{\mathrm{opt}}=1/\gamma_m^{\mathrm{opt}}=1/(\gamma_m^{\mathrm{FB}}+\gamma_m^{\mathrm{COMP}})$ by shining $L_{\mathrm{FB}}$ and $L_{\mathrm{COMP}}$ on the molecules during the hold time between the push pulse and dSTIRAP, at several intensities of $L_{\mathrm{FB,COMP}}$. We derive $\gamma_m^{\mathrm{FB}} = \unit[2\pi \times 11(1)]{Hz / (W/cm^{2}) }$ and $\gamma_m^{\mathrm{COMP}} = \unit[2\pi \times 60(6)]{Hz / (W/cm^{2})}$.

\section{Conclusion and outlook}
\label{sec:conclusion}

In conclusion, we have used STIRAP to associate pairs of Sr atoms in the ground state of lattice sites into weakly-bound ground-state molecules. This association process was efficient despite operating in the regime of strong dissipation $\Omega_{\mathrm{FB}} \lesssim \gamma_{e}$. By making use of a deep optical lattice using 1064-nm light, but without compensating dynamic light shifts created by the STIRAP lasers, we were able to reach a transfer efficiency of roughly $\unit[50]{\%}$. The improvement from $\unit[30]{\%}$ reached in our previous work \cite{Stellmer2012Sr2Mol} is mainly due to the increase in molecule lifetime from $\unit[60]{\mu s}$ up to $\unit[100(20)]{s}$, which was possible by using lattice light of 1064 instead of $\unit[532]{nm}$. The lifetime is now proportional to the tunneling time in the lattice.

The efficiency of the STIRAP scheme without compensation beam is limited by the finite lifetime of the dark state. This lifetime is mainly the result of time-dependent light shifts induced by the free-bound laser on the binding energy of the ground-state molecule. We have identified the coupling to the atomic ${^1S_0} - {^3P_1}$ transition as the main source of light shift, and we have shown how to cancel it with an auxiliary compensation beam, leading to an efficiency higher than $\unit[80]{\%}$. This efficiency is limited in equal parts by scattering from off-resonant light and finite dark-state lifetime. We have shown the effect of the time-independent inhomogeneous lattice light shifts on STIRAP. In further work these detrimental effects could be suppressed by using lattice beams with bigger waists, leading to larger molecular samples.

We thus demonstrated that, by use of STIRAP, we can optically associate atoms into molecules with an efficiency comparable to that obtained with magneto-association through a Feshbach resonance. This general technique can represent a valuable alternative for associating molecules containing non-magnetic atoms, and in particular for the creation of alkali --- alkaline-earth dimers \cite{Hara2011DoubleQDegLiYb, Khramov2014MixGrounLiExcitedYb, Bruni2016PARbYbHyperfineChange, Pasquiou2013RbSrDoubleBEC}. Such dimers have been attracting great attention as they can allow fascinating quantum simulations, thanks to their permanent magnetic and electric dipole moments \cite{Micheli2006ToolboxPolMol}. Finally, coherent, efficient and controlled creation of long-lived ultracold $\mathrm{Sr}_2$ molecules in the ground state could be useful in metrology experiments, for instance as a probe for the time variation of the electron-to-proton mass ratio \cite{Zelevinsky2008MassRatioProtonElectron, Beloy2011VarFundConstMassRatio}.

\vspace{11mm}

\begin{acknowledgments}
We gratefully acknowledge funding from the European Research Council (ERC) under Project No. 615117 QuantStro. B.P. thanks the NWO for funding through Veni grant No. 680-47-438 and C.-C. C. thanks the Ministry of Education of the Republic of China (Taiwan) for a MOE Technologies Incubation Scholarship.
\end{acknowledgments}


%

\end{document}